\documentclass[openacc]{rsproca_new}

\usepackage{bm} 
\usepackage{physics}
\usepackage[english]{babel}

\newcommand{\uv}{\langle uv \rangle}
\newcommand{\av}[1]{\langle {#1} \rangle}
\usepackage{textgreek}

\begin{document}

\title{Conservation laws, fluxes, and symmetries: lessons from a perturbative approach for self-organized turbulence}

\author{
Anna Frishman$^{1}$, S\'ebastien Gom\'e$^{1}$, Anton Svirsky$^{1}$}

\address{$^{1}$Technion Israel Institute of Technology, 32000 Haifa, Israel}

\subject{fluid mechanics, mathematical modelling, statistical physics}

\keywords{two dimensional turbulence, quasi-linear approximation}

\corres{Anna Frishman\\
\email{frishman@technion.ac.il}}

\begin{abstract}
Some turbulent flows self-organize into large-scale structures, rather than breaking up into ever-smaller scales. Underpinning this phenomenon is the existence of two sign-definite quantities which are conserved by the dynamics. Two-dimensional turbulence is a prime example, where large-scale mean flows, termed condensates, spontaneously emerge. We review a perturbative theoretical framework for the statistical description of such inhomogeneous turbulence, offering new perspectives on the role of the two conserved quantities.
We illustrate the universal properties of the theory, comparing results from two-dimensional Navier-Stokes to those from the large-scale-quasi-geostrophic equation. These two models are limiting cases of the shallow water quasi-geostrophic equation, the former exhibiting long-range fluid element interactions, while the latter has local interactions.
We then demonstrate these theoretical ideas in two new settings: first, in rotating three-dimensional turbulence, where two-dimensional condensates are known to form. Considering jet-type condensates, we derive the mean-flow profile and discuss a surprising symmetry breaking. Second, we vary the Rossby deformation radius  in the shallow water quasi-geostrophic equation. We obtain novel domain-spanning condensates in all tested regimes and show that they follow two-dimensional Navier-Stokes for deformation radii above the forcing scale, and the large-scale quasi-geostrophic equation 
for those below, demonstrating the power of these asymptotic models.
 \end{abstract}


\begin{fmtext}

\end{fmtext}
\maketitle

\section{Introduction}
\label{sec:intro}

Paradoxically, turbulence can self-organize.
That is, chaotic small-scale motions can generate large-scale coherent motions. 
This remarkable property can be traced to the presence of two quadratic sign-definite quantities, dynamically conserved at scales where energy injection and dissipation are too slow or absent.
The two-dimensional Navier-Stokes (2DNS) equations are perhaps the simplest example of such a system, serving as the basic model to study such self-organization. In 2DNS, the simultaneous conservation of energy and enstrophy (squared vorticity) implies that they cannot both be transferred to small scales, causing energy to instead flow to large scales. 
In a finite domain and for small enough large-scale dissipation, this energy accumulates at the largest available scale(s), producing a strong mean flow termed a condensate.

A key challenge is to describe statistically this inhomogeneous turbulence, and in particular the emergent mean flows, 
ubiquitous in geophysical, astrophysical and industrial settings \cite{vallis_atmospheric_2017, galperin_zonal_2019,diamond_modern_2010}.
This requires determining the energy exchange between the mean flow and fluctuations, which must be determined self-consistently as the former affect the dynamics of the latter.
We will review the substantial progress made in this direction, emphasizing that an analytic perturbative approach becomes possible, and even soluble, when fluctuations are weak compared to the mean flow (the quasi-linear approximation) and occur at much smaller scales.
%
We will describe how conservation laws, fluxes and symmetries manifest themselves in self-organized turbulence, and demonstrate the power of the perturbative approach for determining and understanding the self-organized steady state.

First, to bring out the universal properties shared by such systems, we will use two model examples, differing by the range of fluid-element interactions:
(1) the two-dimensional Navier-Stokes (2DNS) equations (long-range interactions), and (2) the large-scale quasi-geostrophic (LQG) equation (local interactions). The two arise as distinct asymptotic limits of the same equation, the shallow-water quasi-geostrophic (SWQG) dynamics, 
a basic model for geophysical applications, see \cite{vallis_atmospheric_2017}. 

Both LQG and 2DNS lack an intrinsic temporal or spatial scale. We will next provide a taste of the formalism in settings where such additional scales do exist. We will first venture into the third dimension, considering condensates in rotating three-dimensional fluids. There, the rotation rate provides a temporal scale. We will explain how such condensates fit within the theoretical framework, and describe how the condensate profiles, and their symmetries, change as the temporal scale is varied. To explore the effect of an intrinsic spatial scale we will return to the SWQG equation, varying the range of spatial interactions (the Rossby deformation radius), and measuring the resulting condensates against the asymptotic limits, 2DNS and LQG.


This article is organized as follows. 
We motivate our two study cases from shallow-water quasi-geostrophy in Sec.~\ref{sec:background} and discuss their "anti-diffusive" behavior in statistical steady state in Sec.~\ref{sec:anti-diffusive}. 
In Section~\ref{sec:perturbative_approach_condensate}, we introduce the perturbative description of self-organized turbulence, emphasizing the role of conserved quantities in determining the energy exchanges between fluctuations and mean.
Closure and the resulting condensate profiles are presented in Sec.~\ref{sec:profiles}. 
Second-order statistics, and the role that parity and time-reversal symmetries play in determining them, are discussed in Section \ref{sec:two_point}. 
Throughout, we illustrate the ideas using the prototype of
a uni-directional jet in the two-dimensional Navier-Stokes equation without bottom drag, obtaining its mean flow profile here for the first time.

New material is presented 
in sections \ref{sec:rotating} and \ref{sec:SWQG_vary_Ld}.
In Sec. \ref{sec:rotating}, we derive the mean flow profile of a jet condensate in the rotating 3D Navier-Stokes equations and discuss emergent spatial fluxes and symmetry breaking.
Section \ref{sec:SWQG_vary_Ld} presents novel domain-scale condensates in the shallow-water quasi-geostrophic dynamics. We establish their close similarities with 2D Navier-Stokes and large-scale quasi-geostrophic condensates depending on the Rossby deformation radius.
We summarize and discuss some challenges to the theory in Sec.~\ref{sec:conclusion}.

\section{The shallow-water quasi-geostrophic equation and its limits}
\label{sec:background}
When sufficiently constrained, three dimensional flows become effectively two dimensional. The simplest example is flows at large scales in a thin layer, where the flow in the direction normal to the layer is suppressed. Rotation is another such constraint. In the limit of fast rotation, the dominant force balance is between the Coriolis force and the horizontal pressure gradient, imposing two-dimensional flow. Taking into account inertia as a small deviation from this equilibrium produces dynamics corresponding to the quasi-geostrophic regime. This regime is relevant for large-scale flows in the atmosphere and oceans, which are shaped by rotation and stratification~\cite{vallis_atmospheric_2017}. Idealized models of geophysical flows are thus two dimensional. 

Perhaps the simplest quasi-geostrophic model is that of a shallow layer of fluid under the combined influence of gravity and rotation. Assuming that the layer has an average height given by $H$ (which the free surface fluctuates about) and a constant rotation rate $\Omega \hat{\bm{z}}$, a natural emergent length scale is the Rossby deformation radius $L_d= \sqrt{g_0H}/(2\Omega)$, where $g_0$ is the gravitational acceleration. The quasi-geostrophic regime is obtained in a distinguished limit of fast rotation and strong stratification, see~\cite{vallis_atmospheric_2017}.
In this limit, the horizontal flow is non-divergent at leading order, so that a stream-function $\psi$ can be used for the two-dimensional velocity, $\bm{v}=\bm{\hat{z}}\times\bm{\nabla} \psi$. Hydrostatic balance relates variations in the layers depth, $\delta h$, to the pressure: $g_0 \rho \bm{\nabla} \delta h = \bm{\nabla}p$. Then, geostrophic balance connects the pressure to the stream-function, giving $\psi = (g_0/2\Omega)\delta h$. The SWQG dynamics are thus captured by a single equation for $\psi$:
\begin{align}
    \partial_t q+\bm{v}\cdot \nabla q=g +D + D_{\nu_p} ; \quad q=\left(\nabla^{2}-L_{d}^{-2}\right)\psi ; \quad  \bm{v}=\hat{\bm{z}}\times \bm{\nabla} \psi && \text{SWQG}
    \label{eq:SWQG}
\end{align}
where we have introduced $g$ and $(D+D_{\nu_p})$, the forcing and dissipation, respectively.  This is the Charney-Hasegawa-Mima equation, but here in the absence of differential rotation and thus no Rossby waves (drift waves due to a density gradient in the context of magnetized plasmas), see \cite{diamond_modern_2010,vallis_atmospheric_2017} and reviews in \cite{bouchet2012statistical,galperin_zonal_2019} including waves. 

The quantity $q$ is termed potential-vorticity, satisfying $q=\omega -\psi/L_d^2$ where $\omega =\bm{\nabla} \times \bm{v}$ is the vorticity. It has the dynamics of an active scalar, advected by a velocity which is determined by $q$ itself. Therefore, potential vorticity and all of its moments are Lagrangian, and thus also Eulerian~\cite{falkovich_FluidMechanics_2018}, inviscid conservation laws. In particular, $Z_q\equiv \int q^2/2 ~\dd^2x$ is conserved. 
These conservation laws stem from the conservation of the angular momentum of fluid elements in the vertical direction. For a layer of constant height this leads to the conservation of vorticity, well-known for two-dimensional flows. Here, background rotation and the stretching and compression of fluid columns lead to an additional contribution to the angular momentum per unit mass: $\Omega/(H+\delta h)\approx-\psi/L_d^2$ (at leading order). 
The inviscid system also conserves the total energy, given by the sum of the potential and kinetic energies of the fluid $E=-\int \psi q/2 ~ \dd^2x=\int ((\nabla \psi)^2+\psi^2/L_d^2)/2 ~\dd^2x=\int (\bm{v}^2+\psi^2/L_d^2)/2 ~\dd^2x$. The conservation of potential vorticity gives an idea of how potential energy can be converted to kinetic energy in this system: to keep $q$ unchanged, height variations lead to the generation of vorticity. 

SWQG forms a family of two-dimensional models, depending on the value of the Rossby deformation radius $L_d$. This length-scale corresponds to an effective screening length, beyond which the velocity field induced by a vortex patch decays to zero. It can thus be thought of as a fluid-element interaction range. If $L$ is the box-scale, then the limit $L_d\gg L$ (the limit $\text{Bu}=(L_d/L)^2\gg1$  where Bu is the Burgers number) corresponds to a system with long-range interactions, reproducing 2D Navier-Stokes (2DNS) 
\begin{align}
    \partial_t \omega + \bm{v} \cdot \nabla \omega =g+D + D_{\nu_p}, && \omega =\nabla^2\psi, && \bm{v}=\hat{\bm{z}}\times \bm{\nabla} \psi, && \text{2DNS}
    \label{eq:2DNS}
\end{align}
where non-local interactions arise due to the fluid being incompressible. 
Indeed, changes in the surface height can be neglected in this limit: potential vorticity then reduces to regular vorticity and the energy is dominated by the kinetic energy. For 2DNS, the two quadratic conserved  quantities are the kinetic energy $E_k =\int \bm{v}^2/2 \dd^2x$ and the enstrophy $Z_{\omega}=\int \omega^2 /2 \dd^2x$. 

From this perspective, the opposite limit of very small $L_d$ is also interesting, as it corresponds to a fluid with completely localized fluid element interactions. The relevant limit is then taking $L_d\ll l_f$, in which case 
the SWQG equation \eqref{eq:SWQG}
reduces to the so-called large-scale quasi-geostrophic (LQG) dynamics \cite{larichev_weakly_1991,smith_turbulent_2002}:
\begin{align}
    \label{eq:LQG}
    \partial_{\tau} \psi +\bm{v^{\omega}}\cdot\bm{\nabla}  \psi  =g+D + D_{\nu_p}; && \omega =\nabla^2\psi, && \bm{v^{\omega}}=\hat{\bm{z}}\times \bm{\nabla} \omega, && \text{LQG}
\end{align}
Here we are using a rescaled time $\tau =(L_d/L)^2 t$, as the dynamics in this limit becomes slow, and the streamfunction is redefined to be multiplied by $ L^2$ (as well as terms entering forcing and dissipation).
Note that compared with 2DNS the vorticity and stream-function have here switched roles and that $\bm{v^{\omega}}\cdot\bm{\nabla}  \psi=-\bm{v}\cdot\nabla \omega$. For LQG, the kinetic energy $E_k=\int \bm{v}^2/2 \dd^2x$ and the potential energy $E_p = \int \psi^2/2 \dd^2x$ (up to a proportionality constant) are separately conserved, and form the two quadratic invariants \cite{smith_turbulent_2002}~\footnote{The LQG limit $L_d \ll L $ ($\text{Bu}\ll1$) remains consistent with the quasi-geostrophic limit if $\text{Fr}^2\ll \text{Ro}\ll \text{Fr}$ where $\text{Fr}=U/\sqrt{g_0 H}$ is the Froude number and is a measure of the stratification, the Rossby number  $\text{Ro}=U/(2\Omega L)$ is a measure of rotation, and $U$, and $L$ are characteristic velocity and length scales  (for a detailed derivation see Appendix~A in \cite{svirsky_statistics_2023}).}.

Our focus in this article is on the forced-dissipative dynamics and the corresponding statistical steady state.
The forcing is stochastic, white-in-time and with zero mean, so that it injects 
total (kinetic + potential) energy at fixed mean rate $\epsilon = \langle g \psi \rangle$, which is thus a control parameter.
%
%
Spatial correlations are forced at the characteristic scale $l_f$ throughout.
The turbulent behavior we are interested in emerges when the forcing scale is well separated from dissipative scales, i.e. in the presence of extended inertial ranges.
For small-scale dissipation, we use hyper-viscosity $D_{\nu_p}=-\nu_p (-\nabla^2)^p \omega$ (with $p>2$) in simulations in order to obtain sufficient scale separation between $l_f$ and $l_{\nu_p}$, the corresponding hyper-dissipative scale \footnote{This is common usage in numerical works, in order to obtain sufficient scale separation between $l_f$ and $l_{\nu_p}$, and will not affect our theoretical considerations.}. In addition, we will consider two types of dissipation $D$: 
\begin{align}
    D_\alpha=-\alpha \omega; \quad && D_\nu= \nu\nabla^2 \omega.
\end{align} 
The first, $D_\alpha$, is linear (Ekman) friction, modeling bottom drag, which is the main source of dissipation for a thin layer of fluid. 
When present, friction provides the dominant dissipation mechanism at large scales. The second, $D_\nu$, corresponds to regular viscosity. 
In the absence of friction, viscosity will eventually remove energy at large scales, like the domain scale $L$, with a typical rate $\sim \nu/L^2$.


%
%
%



For 2DNS with two extended inertial ranges $l_{\nu_p}\ll l_f\ll L $, kinetic energy $E_k$ is known to be transferred to large scales, while enstrophy $Z_{\omega}$ flows to small scales. More generally, for a system with two positive-definite quadratic invariants, both stemming from the same dynamical field, and scale separation between forcing and dissipation (providing a range where inviscid invariants are conserved), the invariant with fewer derivatives is expected to be blocked from flowing to small scales, transferred to large scales instead, see \cite{fjortoft1953changes, kraichnan_inertial_1967}. Thus, total energy, $E$, flows to large scales for SWQG (while $Z_q$ flows to small scales), and potential energy, $E_p$, flows to large scales for LQG (while kinetic energy $E_k$ flows to small scales).

\begin{figure}[t!]
    \centering
    \includegraphics[width=0.8\columnwidth]{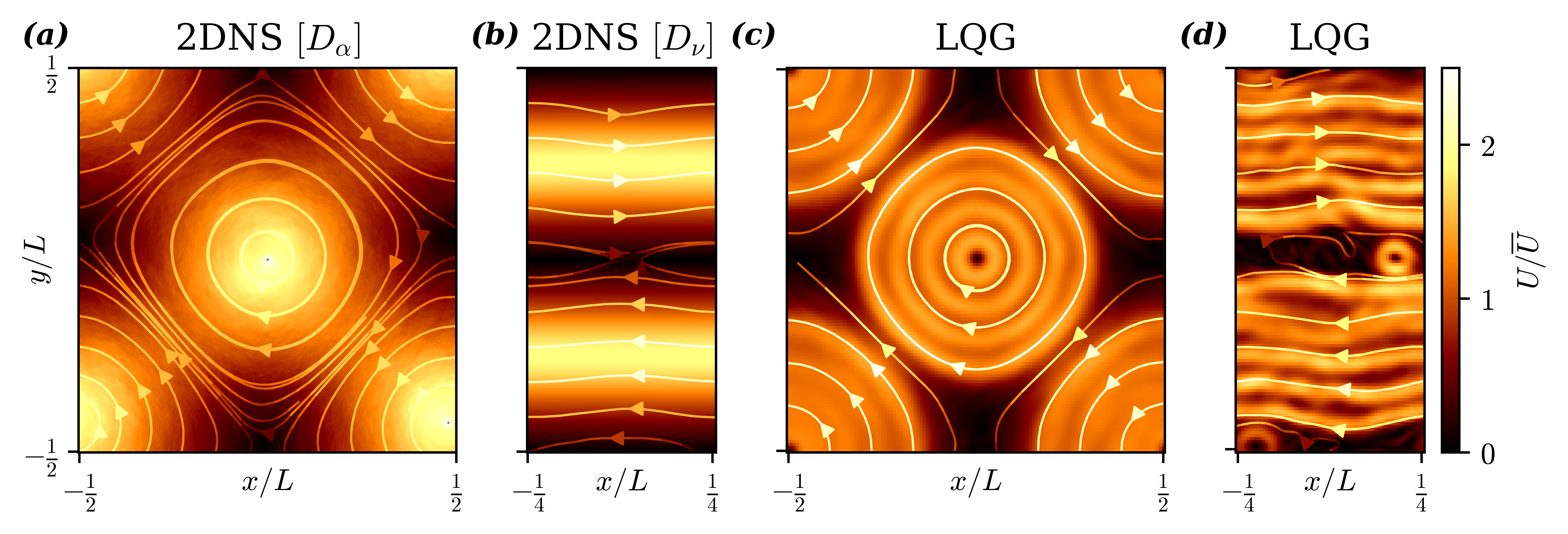}
    \caption{
        Illustration of the vortex and jet condensate configurations from numerical simulations. Shown are snapshots of the velocity field, with color indicating the magnitude, for (a) Vortex condensate 2DNS with friction $(D =D_\alpha)$, (b) Jet condensate of frictionless 2DNS case ($D=D_\nu$); (c)  Vortex condensate in LQG, (d) Jet condensate in LQG.  Data from simulations: (a) \texttt{2DNS(C)}, (b) \texttt{2DNS(D)}, (c) \texttt{LQG(D)}, and (d) \texttt{LQG(B)}, with parameters given in Appendix~\ref{sec:simulation_details}
        }
        \label{fig:vortex_jet}
\end{figure}

In a finite domain, and provided that the energy dissipation rate at the domain scale is slower than the eddy-turnover time at that scale, energy will reach the domain scale and accumulate there. As a result, a large-scale coherent flow, termed a condensate, spontaneously forms, fed by the small-scale fluctuations \cite{kraichnan_inertial_1967,batchelor_computation_1969,smith_bose_1993,dynamics_chertkov_2007,xia_spectrally_2009,boffetta_two-dimensional_nodate}. This emergent large-scale, high-amplitude flow, is evident in any 
instantaneous realization
of the system, see Fig~\ref{fig:vortex_jet}. While this structure tends to fluctuate as a whole,
usually a time-scale separation which allows to define a time-averaged mean flow~\cite{frishman_jets_2017}, which will be our approach in the following.
Our interest here is to characterize such self-organized states: what is the resulting mean flow profiles and the statistics of the fluctuations? Which physical mechanisms determine them?

\section{"Anti-diffusive" behavior of the self-organized state }
\label{sec:anti-diffusive}

For a forcing scale much smaller than the box scale, we expect (most) symmetries that are broken by the forcing to be restored statistically for the condensate. Thus, the two natural cases to consider are an isotropic vortex configuration, where the azimuthal velocity depends on the radius but not the angle, and a jet-type mean flow, where the mean flow is directed along $x$ but varies only along $y$. 
Note that as the condensate spontaneously emerges, it is a-priori unknown which global shape it should take. A simple principle is that the condensate should have the symmetries of the domain, and be consistent with the boundary conditions. Hence, in a doubly-periodic square box the condensate usually takes the shape of a vortex dipole, while it is of the form of two opposing jets pointing along the short direction, in a rectangular box with a sufficiently large aspect ratio~\cite{bouchet_random_2009}~\footnote{While these expectations are confirmed for 2DNS, spontaneous symmetry breaking occurs in LQG \cite{svirsky_out--equilibrium_2025}.}. Such jets and vortices are indeed observed in direct numerical simulations (DNS) in a doubly periodic box (where the total vorticity is zero) depending on the aspect ratio~\cite{bouchet_random_2009,frishman_jets_2017}, as shown in Fig~\ref{fig:vortex_jet}.

\subsection*{Momentum and energy balance in 2DNS}
For simplicity, we will focus on a jet-type mean flow pointing in the $x$ direction, $\langle \bm{v}\rangle=U(y)\hat{x}$, considering the friction-less case, with viscous dissipation $D = D_\nu$, not treated in this geometry before. 
 We begin with the Reynolds averaged equations. They can be derived from the Navier-Stokes equation assuming a statistical steady state and statistical homogeneity in the $x$ direction, and are identical to those in wall bounded flows \cite{pope_turbulent_2000}.
In steady state, the momentum balance in the $x$ direction reads
\begin{align}
    \partial_y\left(\langle uv\rangle-\nu \partial_yU\right)=0,  
    \label{eq:momentum_balance}
\end{align}
where $\bm{v}-U\hat{x}=u\hat{x}+v\hat{y}$ are the components of the fluctuating velocity, and $\langle\cdot\rangle$ is a temporal and spatial average over $x$. This equation implies that the momentum flux in the $y$ direction is constant: $
 J_y \equiv \langle uv\rangle-\nu \partial_yU=const$.
The energy balance for the mean flow is given by
\begin{equation}
    \partial_y\left[U(\langle uv\rangle-\nu \partial_yU)\right]=\langle uv\rangle U' -\nu U'^2,
    \label{eq:jet_mean_energy}
\end{equation}
where $U' (y)\equiv \partial_yU (y)$, while that for the fluctuations reads
\begin{equation}
    \partial_y\left(\langle vp\rangle+\frac{\langle vu^2\rangle+\langle v^3\rangle}{2}\right)=-\langle uv\rangle U'+\epsilon+
    \nu( \langle v\nabla^2v\rangle+\langle  u\nabla^2u\rangle),
    \label{eq:TKE_balance}
\end{equation}
where $p$ here denotes the fluctuating pressure.
 The term $U'\langle uv\rangle$ is responsible for the exchange of energy between the mean flow and fluctuations, where $\langle uv\rangle$ is the (off-diagonal) Reynolds stress. In three-dimensional turbulence, the momentum flux $\langle uv\rangle$ is expected to act against the momentum gradient $\partial_yU$, tending to homogenize momentum \cite{pope_turbulent_2000}. 
This implies $U'\langle uv\rangle<0$, corresponding to the production of turbulence.

On the contrary, in two dimensional turbulence, energy is expected to be extracted \textit{by} the condensate \textit{from} the fluctuations. Thus, we expect that $U'\langle uv\rangle>0$, i.e. the momentum flux tends to  reinforce the mean flow gradient rather than relaxing it, which amounts to an \textit{anti-diffusive} behavior (explicitly so, if $\langle uv\rangle$ were to be modeled to be proportional to the mean flow gradient, see discussions in~\cite{kraichnan_eddy_1976,eyink_turbulent_2006} for homogeneous turbulence). We will show in Sec.~4b\ref{sec:local_2DNS} how such a result can be linked to the conservation of enstrophy, as might be expected, but 
also results in non-trivial space-local energy transfers.

Another distinction from wall-bounded flows where similar equations arise, is that here there is no external inhomogeneity, as the jets spontaneously form (compare to two-dimensional wall-bounded flows ~\cite{falkovich_turbulence_2018} or \cite{nazarenko2000exact} as well). Thus, there is no preferred direction for the momentum flux (independent of the mean flow) meaning that $J_y=0$. A similar conclusion is reached for a frictionless vortex mean flow, when considering the angular momentum flux~\cite{doludenko_coherent_2021}. We thus have
\begin{equation}
    \langle uv\rangle (y) =\nu U' (y).
    \label{eq:mf_local}
\end{equation}
At the same time, as the momentum flux determines the mean-flow energy flux $J_E=U J_y$, see Eq.\eqref{eq:jet_mean_energy}, we get that the mean flow energy balance is local:
\begin{equation}
    \langle uv\rangle U'=\nu U'^2,
\end{equation}
so that all of the energy transferred from the fluctuations is locally dissipated by the mean flow. This is in distinction from the case with linear friction $D=D_\alpha$, where momentum is not conserved and thus does not satisfy a continuity equation. Indeed, it was found that for a vortex condensate with $D=D_\alpha$ the local energy dissipation by the condensate is three times larger than the local energy transfer~\cite{laurie_universal_2014}. 
Further progress requires closure, which is discussed in section~\ref{sec:perturbative_approach_condensate}.

\subsection*{Mass and potential energy balance in LQG}
Similarly to 2DNS, LQG has an additional conserved quantity on top of potential and kinetic energy:
the average fluid height $\int \psi \dd^2x$, reflecting mass conservation. 
We again decompose the flow into the mean $\Psi = \langle \psi \rangle$ and fluctuations $\tilde{\psi} = \psi - \Psi$. Considering the averaged equations with friction ($D= D_\alpha$), and assuming $V_i^{\omega} \partial_i \Psi=0$ with $V_i^{\omega} = \av{v_i^{\omega}}$ --- as is the case for jet or vortex shear flows~\footnote{This holds more generally in regions where the mean flow is strong per the quasi-linear approximation discussed in section~\ref{sec:perturbative_approach_condensate}}, mass conservation reads:
\begin{equation}
\label{eq:ReAvg_LQG}
\partial_{i}\left[
\langle \tilde{v}_{i}^{\omega}\tilde{\psi}\rangle  - \alpha\partial_i \Psi\right]=0,
\end{equation}
where Einstein summation is implied. Here, $\langle \tilde{v}_{i}^{\omega}\tilde{\psi}\rangle  - \alpha\partial_i \Psi$ can be thought of as the total mass flux while $\langle \tilde{v}_{i}^{\omega}\tilde{\psi}\rangle$ is the turbulent mass flux. 
Note the analogy to Eq.~\eqref{eq:momentum_balance}, with the friction term in Eq.~\eqref{eq:ReAvg_LQG} playing the same role as viscous dissipation in Eq.~\eqref{eq:momentum_balance}.
The potential energy balance of the mean flow is then given by:
\begin{equation}
\label{eq:ReAvg_E_MF}
    \partial_i \left(
        \Psi \av{\tilde{v}_i^\omega\tilde{\psi}} - \alpha \Psi \partial_i \Psi
    \right)
    =
    (\partial_i \Psi) \av{\tilde{v}_i^\omega\tilde{\psi}} 
    -\alpha (\partial_i \Psi)^2.
\end{equation}
while the potential energy balance for the fluctuations reads~\footnote{Note that in principle, using the time units in LQG and the redefinition of the stream-function gives that $\langle g(\bm{x},\tau)g(\bm{x'},\tau')\rangle =2\epsilon_p k_f^2  \chi(|\bm{x}-\bm{x'}|) \delta(\tau-\tau')$, where $\epsilon_p=L^4/L_d^2 \epsilon$. However, with a slight abuse of notations, we will denote $\epsilon = \av{g \tilde{\psi}}$ the potential energy injection rate, when discussing LQG.}
\begin{equation}
\label{eq:ReAvg_E_flac}
    \partial_i \left(
        \av{\tilde{v}_i^\omega\frac{\tilde{\psi}^{2}}{2} } 
    \right)
    =
    \epsilon - (\partial_i \Psi) \av{\tilde{v}_i^\omega\tilde{\psi}} +\alpha \langle \tilde{\psi}\nabla^2\tilde{\psi}\rangle.
\end{equation}

The term $(\partial_i \Psi) \av{\tilde{v}_i^\omega\tilde{\psi}}$ is here responsible for the exchange of potential energy between the mean flow and the fluctuations. Potential energy is expected to be transferred \textit{from} the fluctuations \textit{to} the mean flow, corresponding to \textit{anti-diffusive} behavior: $(\partial_i \Psi) \av{\tilde{v}_i^\omega\tilde{\psi}}>0$ meaning that the turbulent mass flux reinforces the mass gradient. In addition, similar symmetry considerations to the ones mentioned for 2DNS imply that the total mass flux must be zero 
and that the potential energy balance for the mean flow is also local in LQG~\cite{svirsky_out--equilibrium_2025}.

\section{A perturbative approach for the condensate}
\label{sec:perturbative_approach_condensate}
Famously, the non-linearities of the Navier-Stokes equation lead to a closure problem in statistical theories of homogeneous and isotropic turbulence, usually making them analytically intractable. Here, instead, the presence of a high amplitude mean flow makes a perturbative approach, termed the quasi-linear (QL) approximation,
feasible, as recognized also in the context of wall-bounded 3D shear flows~\cite{nazarenko2000nonlinear, thomas2015minimal, farrell1993optimal, farrell1993stochastic, pope_turbulent_2000,hwang_attached_2020,marston_recent_2023}. 
In the latter, however, non-linear eddy-eddy interactions play a crucial role in transferring energy to small scales, so they cannot be completely neglected. On the contrary, for systems with an inverse energy transfer such interactions are not strictly necessary, as the transfer can be mediated by the mean flow, so closure is in principle possible. Our focus here is on the statistical steady state, corresponding to the fixed point of the dynamical evolution of the statistics within QL  (formulated using cumulants, called S3T~\cite{Farrell2007,farrell2012dynamics} or DSS~\cite{marston_statistics_2008,Marston2016}), often implemented numerically. 

By itself, however, the quasi-linear approximation is usually insufficient for a tractable closure (e.g. due to the spatially non-local nature of 2DNS, namely the pressure term in Eq.~\ref{eq:TKE_balance}). Moreover the solution is generically expected to depend on the type of forcing, and hence be non-universal. Universality should be restored in the limit of large scale-separation between the forcing and the domain size: $l_f/L\ll1$. Indeed, as we explain below, taking this additional limit produces universal local dynamics, which can be understood in terms of conservation laws and fluxes, facilitating local closure. This approximation underlies several recent successes in deriving mean flow profiles in 2DNS in various settings~\cite{nazarenko2000exact, laurie_universal_2014,doludenko_coherent_2021,van_two_2025}. Here we will add the example of a jet with $D=D_\nu$.   

 \subsection{Assumptions and validity range for QL}
 \label{sec:assumptions}
Within the quasi-linear approximation eddy-eddy interactions are neglected, and energy (potential energy for LQG) is assumed to be transferred non-locally from the fluctuations to the condensate. A classical scale-to-scale inverse cascade is thus assumed to be absent. 
For 2DNS, the validity condition for this approximation is $U'\tau_{e}(k_f)\gg 1$, where $U'$is the mean flow shear-rate and $\tau_e(k_f)$ is the eddy turnover time at the forcing scale (this is also the relevant time-scale at smaller scales, set by the enstrophy flux $\epsilon/k_f^2$ rather than the energy flux). In particular, for 2DNS $1/\tau_e(k_f)\sim \epsilon^{1/3}k_f^{2/3}$ and an estimate for the mean flow shear can be obtained using the energy balance. Assuming most of the injected energy is dissipated by the mean flow, the global energy balance reads $\alpha U^2=\epsilon$ in the presence of friction, or $\nu (U')^2=\epsilon$ if $D=D_{\nu}$. The condensation process implies that energy piles up at the box scale, so that roughly $U'\sim U/L$ and this gives either $U'\sim\sqrt{\epsilon/(\alpha L^2)}$ ($D=D_\alpha$) or $U'\sim \sqrt{\epsilon/\nu}$ ($D=D_\nu$). The condition for the QL approximation to be valid throughout the domain is thus
\begin{align}
    (L/l_f)^{2/3} \delta^{1/2}\ll1, \qquad D=D_\alpha; && Re^{-1/2}\ll1, \qquad D=D_{\nu}
\end{align}
where $\delta \equiv \alpha/\tau_e(L)$ and $Re=\tau_e(k_f)/(\nu k_f^2)$. 
Note that $\delta \ll1 $ is a necessary condition for the condensation to occur for $D=D_{\alpha}$ while for $D=D_{\nu}$ a condensate is guaranteed to form. 
Similarly, the QL condition in the former case is not guaranteed, requiring a particular ordering of the limits $\delta\ll1$ and $l_f/L\ll1$. However, even if it is not satisfied globally, as in~\cite{laurie_universal_2014,frishman_turbulence_2018}, it could still be satisfied locally, depending on the local shear rate. Indeed, for a vortex mean flow the local mean shear is $r\partial_r(U/r)\sim U/r$ so the QL approximation can still be applied for small enough radii $r/L\ll (l_f/L)^{2/3} \delta^{-1/2}$, see discussion in~\cite{kolokolov_structure_2016} and~\cite{frishman_turbulence_2018}. Similar estimates, comparing the typical mean flow time-scale to a non-linear one can be performed for LQG~\cite{svirsky_two_2023}.

\subsection{The role of the additional conserved quantity}
The most fundamental question to be addressed by any closure is: how much energy is exchanged between the mean flow and the fluctuations locally? For 2DNS it has been shown in previous works that within QL, and assuming scale separation $l_f/L\ll1$, all the locally injected energy is transferred to the mean flow, so the local exchange term is exactly equal to the local energy injection rate~\footnote{There are some caveats to this statement, in particular with regards to the order of limits, which we will comment about below.}, i.e.\ of the form:
\begin{equation}
    U' (y) \langle uv\rangle (y)=\epsilon .
    \label{eq:closure}
\end{equation} 
This closure was probably first derived in Ref.~\cite{nazarenko2000exact}. Independently, its relevance for condensates in 2D turbulence was realized in Ref.~\cite{kolokolov_structure_2016}, following the derivation of the condensate profile  in~\cite{laurie_universal_2014}. Before showing the consequences of this result for closure, we here show that it stems from the conservation of enstrophy for the fluctuations. Our analysis will make explicit the interplay between inverse and direct transfers, and the crucial role the latter plays in shaping the former both in 2DNS and in LQG.

\subsubsection{Enstrophy conservation for fluctuations in 2DNS}

\label{sec:local_2DNS}

We start with the quasi-linear equation for the vorticity fluctuations in \eqref{eq:2DNS},
\begin{align}
    \partial_t \omega + U(y) \partial_x \omega  - v \partial_y^2 U(y) = g+ D.
    \label{eq:QL_vorticity}
\end{align}
In the presence of scale separation between the mean flow and fluctuations, the term $-v \partial_y^2 U$ in Eq.~\eqref{eq:QL_vorticity}, can be neglected. Then, enstrophy is conserved separately for the fluctuations. Indeed, the exchange of enstrophy between the mean flow and the fluctuations, $|(\partial_y^2U) \tilde{\omega}\partial_x\tilde{\psi}|$, where $\tilde{\omega}\equiv\omega-\langle\omega\rangle$, becomes slow in this limit:  $|(\partial_y^2U) \tilde{\omega}\partial_x\tilde{\psi}|/ |U(y)\partial_x \tilde{\omega}^2/2|\sim (l_f/L)^2\ll1$, and $\int \tilde{\omega}^2/2 \dd^2x$ is conserved by the dynamics. While this seems like a trivial fact, it implies the fluctuations enstrophy equation takes a conservative form, with an associated local spectral flux of enstrophy.
 
We begin with the equation for the vorticity two-point function, considering small separations such that we can expand $U(y_2)\sim U(y_1)+U'(y_1)(y_2-y_1)$ under the scale separation assumption, where $U'\equiv\partial_y U(y)$, to obtain:
\begin{align}
    \partial_t \langle \tilde{\omega}(\bm{x}_1,t)\tilde{\omega}(\bm{x}_2,t)\rangle+U'(y_1)(y_1-y_2)\partial_{x_1} \langle \tilde{\omega}(\bm{x}_1,t)\tilde{\omega}(\bm{x_2},t)\rangle=2\epsilon k_f^2 \chi(\bm{x}_1-\bm{x}_2)+D(\bm{x}_1,\bm{x}_2)
    \label{eq:correlations2D}
\end{align}
with $D(\bm{x}_1,\bm{x}_2)$ a small-scale dissipation term (e.g. $D(\bm{x}_1,\bm{x}_2)= \nu(\langle \omega (\bm{x}_1)\nabla^2 \omega (\bm{x}_2) \rangle +  \langle \omega (\bm{x}_2)\nabla^2 \omega (\bm{x}_1) \rangle )$ for regular viscosity)
and $\chi (\bm{x})$ the forcing two-point correlation function, such that $\langle g(x,t)g(x',t)\rangle =2\epsilon k_f^2  \chi(x-x') \delta(\tau-\tau')$ and $\chi(0) = 1$.
Note that we have used homogeneity in $x$ (implying that $\partial_{x_1}=-\partial_{x_2}$) to obtain Eq.~\eqref{eq:correlations2D}.

Following \eqref{eq:correlations2D}, the fluctuations in a neighborhood of the point $y_1$ will effectively feel a linear mean shear of magnitude $U'(y_1)$, which will then facilitate the closure~\cite{kolokolov_structure_2016}. Indeed, the evolution of modes in the background of a linear shear is a classical and analytically tractable problem, going back to Lord Kelvin. It has also been considered as a solvable example for jet dynamics on the beta plane (assuming a linear profile a-priori)~\cite{Srinivasan2012,srinivasan_reynolds_2014,bouchet_kinetic_2013}.
Correlation equations such as \eqref{eq:correlations2D} were previously discussed in the context of $\beta-$plane jets and drift-wave plasma turbulence, e.g.~\cite{Srinivasan2012, srinivasan_reynolds_2014, parker_dynamics_2016, parker2013zonal, parker2014generation, jackman_parameterisation_2023}. 
Here, assuming scale separation, Eq.~\eqref{eq:correlations2D} allows a \textit{self-consistent} closure for the mean flow, giving the energy transfer between the mean flow and the fluctuations as a function of $U'(y_1)$\cite{nazarenko2000exact, laurie_universal_2014,kolokolov_structure_2016,frishman_culmination_2017,jackman_parameterisation_2023}. Rather than repeating these calculations here, we will offer a spectral perspective on these results.  

Defining $\mathcal{Z}(\bm{k},y_1)$ to be the Fourier transform of $\langle \tilde{\omega}(\bm{x}_1,t)\tilde{\omega}(\bm{x_2},t)\rangle/2$ with respect to the variables $\Delta x =x_1-x_2, \Delta y =y_1-y_2$, such that $(\Delta x, \Delta y)\to (k_x,k_y)$,
we obtain:
\begin{equation}
    \partial_t \mathcal{Z}(\bm{k},y_1)+\partial_{k_y}\left(-U'(y_1)k_x\mathcal{Z}(\bm{k},y_1)\right)=\epsilon k_f^2 \chi_{\bm{k}}+\mathcal{D}(\bm{k},y_1)
    \label{eq:enstrophy_flux}
\end{equation}
where $\mathcal{D}(\bm{k},y_1)$ and $\chi_{\bm{k}}$ are the corresponding Fourier transforms of the dissipation and forcing terms and we keep an explicit dependence on $y_1$. We thus arrive at a continuity equation in spectral space, where $\Pi^Z(\bm{k},y_1)=-U'(y_1)k_x\mathcal{Z}(\bm{k},y_1)$ is the flux of (the fluctuations) enstrophy along $k_y$ (note that $k_x$ remains unchanged under the effective shear dynamics, and is equal to its value at the forcing scale). In steady state, at scales $k_y$ unaffected by forcing and dissipation, we expect a constant flux of enstrophy to develop, equal to the enstrophy injection rate.
In particular, in the inertial range, 
\begin{align}
&k_f\ll |k_y|\ll k_d: \\
    &|U'(y_1)k_x|\mathcal{Z}(k_x,k_y,y_1)=k_f^2\epsilon_{k_x},  &&\quad \text{if} \qquad  k_yk_xU'(y_1)<0 \nonumber \\
    &\mathcal{Z}(k_x,k_y,y_1)=0 &&\quad \text{otherwise}    
    \label{eq:flux}
\end{align}
where $k_d$ is a dissipative scale (to be specified in the following), 
and $k_f^2\epsilon_{k_x}=\int k_f^2\epsilon\chi_{\bm{k}}dk_y$ is the total enstrophy injection rate for modes with the given $k_x$ (assumed to be peaked around $k_x^2+k_y^2=k_f^2$). The quadrants $(k_x,k_y)$ in which the flux is non-zero can be deduced from the solution of the dynamical equation \eqref{eq:energy_balance} along characteristics $(k_x, k_y-k_x U'\tau)$: enstrophy $\mathcal{Z}(\bm{k}, y_1)$ injected at the forcing scale is advected towards positive $k_y\to \infty$ if $U' (y_1) k_x <0$, or negative $k_y\to -\infty$ otherwise.

In agreement with the phenomenology of 2D turbulence~\cite{kraichnan_inertial_1967}, the existence of a flux of enstrophy to small scales limits the small scale flux of the fluctuations energy. Indeed, consider $\mathcal{E}(\bm{k},y_1)=\frac{\mathcal{Z}(\bm{k},y_1)}{k^2}$, the corresponding fluctuations energy, equal at leading order in $l_f/L$ to the Fourier transform of $\langle \nabla\tilde{\psi}(\bm{x}_1) \cdot \nabla\tilde{\psi}(\bm{x}_2)\rangle$ (keeping derivatives only with respect to the difference coordinates $(\Delta x,\Delta y)$, as in~\cite{galperin_statistical_2019,parker_dynamics_2016}). 
Then equation \eqref{eq:enstrophy_flux} can be rewritten as:
\begin{equation}
        \partial_t \mathcal{E}(\bm{k},y_1)+\partial_{k_y}\left(-U'(y_1)k_x\mathcal{E}(\bm{k},y_1)\right)= 2U'\frac{k_xk_y}{k^2}\mathcal{E}(\bm{k},y_1)+ \epsilon  \chi_{\bm{k}}+\mathcal{D_E}(\bm{k},y_1)
        \label{eq:energy_balance}
\end{equation}
where $\mathcal{D_E}=\mathcal{D}/k^2$. While the fluctuations enstrophy is separately conserved, the energy is not.
Indeed, the terms in Eq.~\eqref{eq:energy_balance} have a familiar interpretation: $\Pi^E=-U'(y_1)k_x\mathcal{E}(\bm{k},y_1)$ is the spectral energy flux in the $k_y$ direction, while $2 U'\frac{k_xk_y}{k^2}\mathcal{E}(\bm{k},y_1)$ is the contribution of mode $(k_x,k_y)$ to the energy exchange between the mean flow and fluctuations (i.e. to $- U' \langle uv\rangle$).
Note that only modes with $-k_f\lesssim k_x\lesssim k_f$, initially excited, give a non-zero contribution. In steady state, using Eq.~\eqref{eq:flux}, we find that the energy flux in the inertial range is given by 
\begin{align}
k_f\ll |k_y|\ll k_d: \qquad 
     \Pi^E(k_x,k_y,y_1)=U'(y_1)k_x\frac{\mathcal{Z}(\bm{k},y_1)}{k^2}=\frac{k_f^2}{k_x^2+k_y^2}\epsilon_{k_x} 
     \label{eq:energy_flux}
\end{align}
here written for $k_y>0$ and $k_xU'(y_1)<0$. Thus, the flux is positive but asymptotically vanishing as $k_y\to\infty$. Physically, shearing by the mean flow generically causes the transfer of the fluctuations energy to small scales. This also generates a flux of enstrophy to small scales, and the conservation of the latter implies the fluctuations energy flux must vanish at small scales. Then, the energy must either be transferred to large-scale fluctuations, or to the condensate, via the exchange term on the right-hand-side of equation \eqref{eq:energy_balance}. The former does not occur if the shear can act efficiently on all the excited fluctuations modes, transferring the energy to sufficiently 
large $k_y$, in which case we get that all of the energy is transferred to the condensate. 
Indeed, integrating Eq.~\eqref{eq:energy_balance} over $k_x \in \mathbb{R}$ and $-k_d < k_y < k_d$, we obtain:
\begin{equation}
\begin{split}
    \int_{-\infty}^{\infty} \frac{dk_x}{(2\pi)^2}\left(\Pi^E(k_x,k_y\to k_d)-\Pi^E(k_x,k_y\to-k_d\right) \approx 0 \\
    =U'\int_{-\infty}^{\infty} \frac{dk_x}{2\pi}\int_{-k_d}^{k_d} \frac{dk_y}{2\pi}\frac{k_xk_y}{k^2}2\mathcal{E}(\bm{k},y_1)+\epsilon\approx -U' (y_1) \langle uv\rangle+\epsilon,
    \end{split}
\end{equation}
where in the first line we used the expression for the energy flux \eqref{eq:energy_flux}, and in the last approximate equality we have taken the limit $k_d\to\infty$ and used the expression for the Reynolds stress in Fourier space. This establishes the local energy balance \eqref{eq:closure}.

The above calculation relies on the assumption that $k_d\gg k_f$ (so that enstrophy reaches arbitrarily small scales). An estimate for $k_d$ in the frictionless case is $k_d\sim\sqrt{U'/\nu}$ (changed to $k_d\sim(U'/\nu_p)^{1/2p}$ when using hyper-viscosity, as we do in our simulations), and $k_d$ can be made arbitrarily large for large enough (hyper-viscous) Reynolds number. The deviations from the balance Eq.~\eqref{eq:closure} are then small, suppressed by powers of $k_f/k_d$, as shown in~\cite{kolokolov_structure_2016}.

On the other hand, if $D=D_\alpha$ and linear friction is the main dissipation mechanism at the forcing scale ($\Gamma = \nu_p k_f^{2p}/\alpha \ll 1$), the cutoff $k_d$ is determined by the temporal cutoff $1/\alpha$, and becomes $k_x$-dependent. A more subtle discussion is required then: for energy to be able to reach small-enough scales $k_d \gg k_f$ for all modes $k_x$, the shear must be fast enough compared to the friction at the forcing scale, $U'/\alpha \ll L/l_f$ ($\delta^{1/2}(L/l_f)\ll1$). Otherwise, modes with $k_x\sim 2\pi/L$ extract a large fraction of energy from the condensate. For an infinite domain $L/l_f\to \infty$, this results in the vanishing of the total Reynolds stress $\uv$ \footnote{This can be connected to a Lagrangian conservation law~\cite{frishman_statistical_2015}, similar to the discussion in~\cite{kraichnan_eddy_1976}}, in which case a condensate cannot be sustained, see
\cite{Srinivasan2012,srinivasan_reynolds_2014,kolokolov_structure_2016}. Such exact cancellation is not expected in any finite domain~\cite{frishman_culmination_2017,jackman_parameterisation_2023}, as confirmed with numerical simulations (see \cite{SM}) where we find essentially the same condensate, including the mean flow profile in the QL region, when comparing $\Gamma=10^2$ and $\Gamma=10^{-4}$ simulations \texttt{2DNS(A-B)}, see \cite{SM} for additional details).

The upshot of this section is the result in Eq.~\eqref{eq:closure} 
which will be used below to obtain the mean flow profile in our jet example.

\subsubsection{Kinetic energy conservation for fluctuations in LQG}
\label{sec:KE_LQG}
Owing to the local nature of LQG, it suffices to apply the quasi-linear approximation and neglect dissipation in the potential energy balance \eqref{eq:ReAvg_E_flac} to obtain closure: all the potential energy is then locally transferred to the mean flow \cite{svirsky_statistics_2023}, 
\begin{equation}
    (\partial_i \Psi) (\bm{x} ) \av{\tilde{v}_i^\omega\tilde{\psi}} (\bm{x}) = \epsilon.
    \label{eq:closure_LQG}
\end{equation} 
 However, the conservation of kinetic energy in the scale-separated limit plays an important role in shaping the condensate in LQG.
First, we note that no kinetic energy is exchanged between the mean flow and the fluctuations if $\nabla\Psi=const$, while the equation for the kinetic energy does not become trivial in this limit:

\begin{align}
    \partial_\tau \frac{\langle (\bm{\nabla} \tilde{\psi})^2\rangle}2+U(y) \partial_y\langle \bm{\nabla}\tilde{\psi} \cdot\partial_x\partial_y\bm{\nabla}\tilde{\psi}\rangle =\epsilon k_f^2+  D(\nabla\tilde{\psi}),
\end{align}
written here for a jet mean flow $U(y)=-\partial_y\Psi$, neglecting cubic non-nonlinearities and assuming statistical homogeneity along $x$ and where $D(\nabla\tilde{\psi})$ is a dissipative term. Indeed, expanding $U(y)\approx U$, assumed independent on $y$ at leading order, gives a continuity equation for the kinetic energy where $\Pi^{KE}=U\langle \bm{\nabla}\tilde{\psi} \cdot \partial_x\partial_y \bm{\nabla}\tilde{\psi}\rangle$ is the corresponding spatial flux.

Considering the equation for the two-point function under the same assumptions gives
\begin{align}
    \partial_\tau \langle \bm{\nabla}_1 \tilde{\psi}(\bm{x_1},t) \cdot \bm{\nabla}_2\tilde{\psi}(\bm{x_2},t)\rangle+ U(\nabla_1^2-\nabla_2^2)\partial_{x_1} &\langle \bm{\nabla}_1\tilde{\psi}(\bm{x_1},t)\cdot\bm{\nabla}_2\tilde{\psi}(\bm{x_2},t)\rangle \nonumber\\ &=2\epsilon k_f^2 \chi(\bm{x_1}-\bm{x_2})+D(\bm{x_1},\bm{x_2}),
\end{align}
where $\bm{\nabla}_1$ and $\bm{\nabla}_2$ are gradients with respect to $\bm{x}_1$ and $\bm{x}_2$, respectively, and $D(\bm{x_1},\bm{x_2})$ is a dissipation term due to friction and hyper-viscosity. 
The equation for the two-point function, while retaining a coupling between the condensate and the fluctuations, has constant coefficients, independent of $y$. Thus, the condensate does not couple modes at different scales (different wavenumbers), at leading order in $l_f/L$. In particular, there is no energy flux to small scales where the condensate dominates (hence where QL applies). Then, since the kinetic energy of the fluctuations cannot be transferred to the condensate, and cannot be dissipated at small scales (as it does not reach such scales), the remaining option for a steady state is the expulsion of the kinetic energy from the region of the condensate, i.e. the formation of a spatial flux of kinetic energy away from the condensate:
\begin{equation}
    U(y)\partial_y\langle\nabla_i\tilde{\psi} \partial_x \partial_y \nabla_i\tilde{\psi}\rangle=\epsilon k_f^2.
\end{equation}
This relation and the corresponding flux was explicitly computed in~\cite{svirsky_two_2023}.

\begin{figure}[t!]
    \centering
    \includegraphics[width=1\columnwidth]{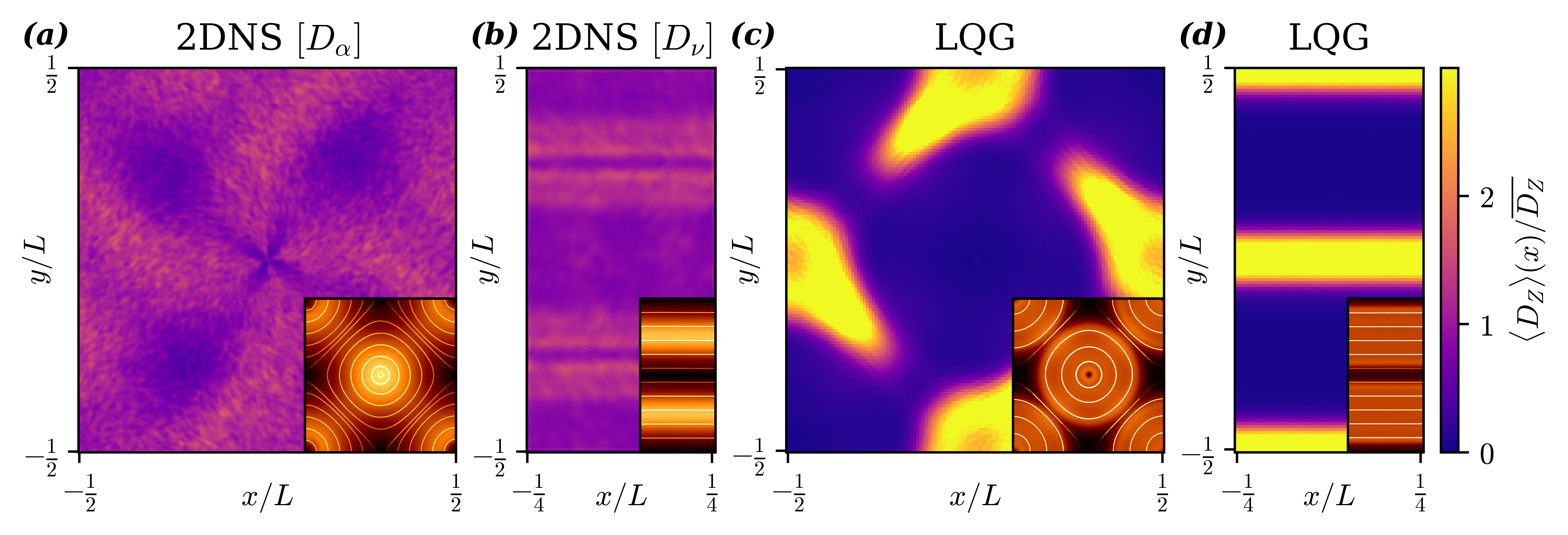}
    \caption{Time-averaged small-scale dissipation $\av{D_Z}(x)$ of (a,b) enstrophy in 2DNS and (c,d) kinetic energy in LQG, normalized by the spatial mean $\overline{D_Z}$. Insets show the corresponding mean velocity fields for reference. Simulations shown: (a) \texttt{2DNS(C)}, (b) \texttt{2DNS(D)}, (c) \texttt{LQG(D)}, and (d) \texttt{LQG(B)}.
    }
    \label{fig:diss_map}
\end{figure}

In LQG, there is thus a partitioning of space between a condensate-dominated region where an inverse transfer of potential energy occurs, and a second region, where the condensate amplitude is very small, thus outside of the QL approximation, and a direct cascade of kinetic energy occurs. Scales smaller than the forcing are only generated in the latter regions, and that is also where small-scale 
dissipation dominates. This can be clearly seen in Fig~\ref{fig:diss_map}, 
where the small-scale enstrophy dissipation for 2DNS 
and kinetic energy dissipation for LQG 
are shown both for a vortex dipole and for an alternating-jet condensate. 
While the dissipation of enstrophy for 2DNS is mostly homogeneous, 
implying that the direct transfer of enstrophy to small scales is essentially unaffected by the inverse transfer, the LQG simulations have all the small-scale dissipation of kinetic energy concentrated in regions where the condensate amplitude is suppressed. 

\section{Closure for the mean flow profile}
\label{sec:profiles}
\begin{figure}[t!]
    \centering
    \includegraphics[width=1\columnwidth]{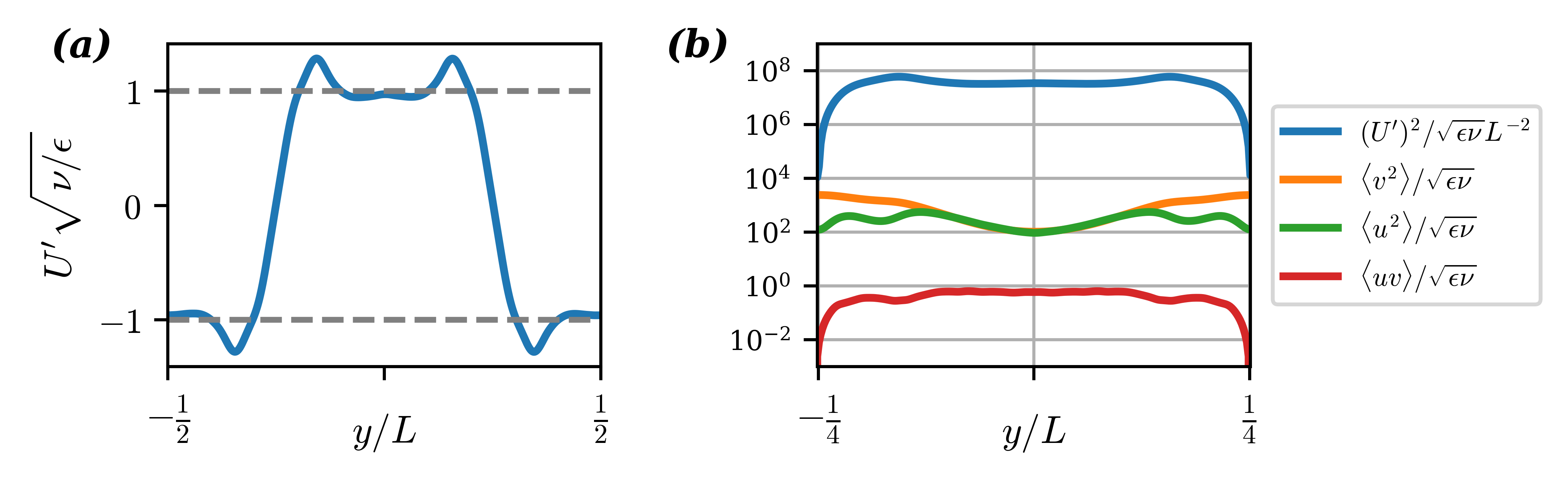}
    \caption{Condensate in frictionless 2DNS ($D = D_\nu$). (a) Mean-flow shear rate $U'=\partial_y U(y) $ ; the dashed line corresponds to the closed solution \eqref{eq:jet_nu}. (b) Mean enstrophy $U'^2$ (blue), Reynolds stress $\av{uv}$ and velocity fluctuations $\av{v^2}$ (orange) and $\av{u^2}$ (green). For clarity only a single jet is show (corresponding to $-1/4<y<1/4$ in (a). Numerical data correspond to simulation \texttt{2DNS(D)}.
    }
    \label{fig:2DNS_nu}
\end{figure}

For 2DNS, we now wish to use the quasi-linear approximation and the local result \eqref{eq:closure} to obtain the global profile of the condensate. Note that the type of large-scale dissipation does not alter the closure Eq.\eqref{eq:closure}, but does influence the resulting mean flow profile. For the frictionless jet example, we use the expression for the Reynolds stress in terms of the mean flow shear, Eq.~\eqref{eq:closure}, and the momentum balance, Eq.~\eqref{eq:momentum_balance} to obtain 
\begin{align}
    U'(y)=\pm\sqrt{\epsilon/\nu} && \langle uv\rangle =\pm\sqrt{\epsilon\nu}, 
    \label{eq:jet_nu}
\end{align}
so the emergent jet profile is a linear shear! Of course, a periodic box contains two oppositely-directed jets. The solution in Eqs.~\eqref{eq:jet_nu} describes the regions of monotonic shear of either sign, as confirmed in DNS, see Fig.~\ref{fig:2DNS_nu}(a) for $U'(y)$ and Fig.~\ref{fig:2DNS_nu}(b) for $\uv(y)$ (red line). However, it does not apply at the transition between the two, where $U'=0$ so Eq.~\eqref{eq:closure} cannot apply. This region, where the local expansion of $U(y)$ is inapplicable, is apparently small, as can be seen in Fig.~\ref{fig:2DNS_nu}(a). 
With friction $D=D_\alpha$, the resulting profile for a jet is more complicated and the condensate is observed to consist of both large-scale vortices and jets~\cite{frishman_jets_2017,frishman_culmination_2017}. %

For a vortex-type condensate, the relation \eqref{eq:closure} is replaced by $\langle v_{\phi}v_r\rangle r\partial_r \left(\frac{V_{\phi}}{r} \right)=\epsilon $,
where $V_{\phi}(r)$ is the mean azimuthal velocity~\cite{kolokolov_structure_2016}. 
This leads to the friction-based profile, $|V_\phi|=  \sqrt{3\epsilon/\alpha}$ \cite{laurie_universal_2014}, or the profile $|V_\phi|=\sqrt{\epsilon/\nu} r\ln(r/R)$ for $l_f\ll r\ll R$ without friction \cite{doludenko_coherent_2021}, verified in \cite{van_two_2025}.


In LQG, combining \eqref{eq:ReAvg_LQG} with \eqref{eq:closure_LQG} gives that 
$|U| \equiv |\nabla_i\Psi|= {\rm const}$, and: 
\begin{align}
    \bm{\nabla}\Psi = \sqrt{\frac{\epsilon}{\alpha}} \hat{q_1}; \quad &&
    \bm{U} = \hat{z} \times \bm{\nabla}\Psi =  \sqrt{\frac{\epsilon}{\alpha}} \hat{q_2}; \quad &&
    \av{\bm{\tilde{v}}^{\omega}\tilde{\psi}} &= \sqrt{\alpha \epsilon} \hat{q_1}; 
\end{align}
where $\hat{q_1}, \hat{q_2}$ correspond to an arbitrary basis defining the condensate geometry. The results for a straight jet are obtained by imposing that $(\hat{q_1},\hat{q_2}) = (\hat{x},\hat{y})$, and for a circular dipole condensate by taking $(\hat{q_1},\hat{q_2}) = (\hat{r},\hat{\theta})$, imposing a polar symmetry, confirmed in DNS\cite{svirsky_two_2023,svirsky_out--equilibrium_2025}. 

\section{Two-point correlation functions and PT symmetry}
\label{sec:two_point}
So far, we were able to determine the mean flow profile and the associated off-diagonal Reynolds stress. What about the diagonal parts of the Reynolds stress, $\langle v^2\rangle$, $\langle u^2\rangle$ which determine the total fluctuations energy? Importantly, these moments have rather different symmetry properties than those of $\langle uv\rangle$. The latter is tightly connected to the energy transfer, hence to the breaking of the time reversal $t\to-t$ symmetry. While $\uv$ is invariant under this symmetry, it does break the combined parity (P) $x\to-x$ plus time reversal (T) $t\to -t$ symmetry, transforming as $\langle uv\rangle\to -\langle uv\rangle$. On the other hand, $\langle v^2\rangle$ and $\langle u^2\rangle$ are invariant with respect to PT symmetry.   

More broadly,
consider the equation for the two-point, single time, correlation function
$\langle \tilde{\psi}_1 \tilde{\psi}_2\rangle\equiv \langle \tilde{\psi}(\bm{x}_1,t) \tilde{\psi}(\bm{x}_2,t)\rangle $ for 2DNSE within the QL approximation:
\begin{align}
    &\mathcal{L}^{1,2}_U \langle \tilde{\psi}_1 \tilde{\psi}_2\rangle = 2\epsilon k_f^2\chi(\bm{x}_1-\bm{x_2)}+D(\bm{x_1,\bm{x_2})} \label{eq:two_points} \\
&\mathcal{L}^{1,2}_U=U(y_1)\nabla_{1}^{2}\nabla_{2}^{2}\partial_{x_{1}}+U(y_2)\nabla_{1}^{2}\nabla_{2}^{2}\partial_{x_{2}} 
    =\sqrt{\frac{\epsilon}{\nu}}(y_1-y_2)(\partial_{y_1}^2+\partial_{x_{1}}^2)(\partial_{y_2}^2+\partial_{x_{1}}^2)\partial_{x_{1}} \nonumber
\end{align}
where in the last line we have used the linear shear profile obtained for a jet condensate with $D=D_{\nu}$ to be specific. The operator $\mathcal{L}^{1,2}_U$ is anti-symmetric with respect to PT (since $U\to U$ under this symmetry and $\partial_x\to -\partial_x$), while the forcing and the dissipation correlators are invariant with respect to it. This is just a reflection of the fact that the Euler equation is invariant with respect to this symmetry, while the forcing and dissipation break it.

Equation \eqref{eq:two_points} can be considered at different scales. At the forcing scale and below, but above the dissipative scale so that $D$ can be neglected, it reduces to the local version discussed in Sec.~4b\ref{sec:local_2DNS}. The momentum flux $\langle uv\rangle$ is determined by the forcing, thus stemming from the particular  solution to the inhomogeneous equation \eqref{eq:two_points} with $D$ set to zero. 

On the other hand, we expect most of the energy to be contained at large scales, so that the main contribution to $\langle v^2\rangle$, $\langle u^2\rangle$ comes from scales $|\bm{x_1}-\bm{x_2}|\gg l_f$ where the forcing correlator can be neglected. This suggests that $\langle v^2\rangle$, $\langle u^2\rangle$ are to be determined by the zero modes of the operator $\mathcal{L}^{1,2}_U$ at leading order. 
That such zero modes determine PT invariant second-order moments has been demonstrated for 2DNS~\cite{frishman_turbulence_2018} and LQG~\cite{svirsky_statistics_2023,svirsky_two_2023}. Determining which zero modes are chosen from first principles requires going further in the perturbative expansion, including dissipative terms, non-linear terms, and accounting for boundary conditions (or a combination of all three) and is yet to be done.

We will not attempt to obtain the relevant zero modes for 
the viscous jet condensate \eqref{eq:jet_nu}. Instead, we present evidence that $\langle u^2\rangle$ and $\langle v^2\rangle$ do not stem from the particular solution to equation \eqref{eq:two_points}. This implies that at leading order they should be determined by zero modes, solutions of the homogeneous Eq.~\eqref{eq:two_points}. In particular, for a linear shear, the operator $\mathcal{L}^{1,2}_U$ is invariant with respect to space translations in the $y$ direction. For statistically homogeneous forcing the resulting single point statistics related to the particular solution would thus also be homogeneous, i.e. independent of $y$. This solution was essentially analyzed in Sec.~4b\ref{sec:local_2DNS} since the profile here is a linear shear.
, see also~\cite{srinivasan_reynolds_2014,kolokolov_velocity_2016,jackman_parameterisation_2023}. On the contrary, we find that $\langle v^2\rangle$, $\langle u^2\rangle$ vary by an order of magnitude in a region where the jet mean flow profile remains extremely flat, Fig~\ref{fig:2DNS_nu}, so they do not correspond to these particular solutions. Qualitatively, we find QL is well satisfied in the region where the shear is flat ($\langle u^2\rangle/U^2, \langle v^2\rangle/U^2\ll 1$), and $\langle u^2\rangle\sim \langle v^2\rangle \gg \langle uv \rangle$
%
The latter hierarchy is observed in other settings in 2DNS and LQG: PT-breaking correlators are suppressed compared with PT-invariant ones~\cite{laurie_universal_2014,svirsky_statistics_2023}.        

This suggests a broader picture for the second-order statistics: large-scale, PT-invariant moments are determined by homogeneous solutions to the quasi-linear equation (zero modes of the corresponding operator $\mathcal{L}^{1,2}_U$). On the other hand, PT-breaking correlators arise from the particular solution to the equation, wherein the relevant time-reversal breaking terms are taken into account. Beyond self-organized turbulence, such a structure may be relevant for wall-bounded flows, where PT is not broken by external forcing of the small scale turbulence, but rather by non-linear terms responsible for eddy-eddy interactions which eventually take away the energy to smaller scales (which are however sometimes modeled as effective forcing and dissipation~\cite{hwang_attached_2020}).   

\section{Rotating 3D turbulence: condensate profile}
\label{sec:rotating}
The inverse transfer of energy, and the consequent condensation of energy at the largest scale, are known to occur even in three-dimensional settings which do not have a thin direction, see\cite{alexakis_cascades_2018} and \cite{rubio2014upscale, favier2014inverse, guervilly_jets_2017, guzman2020competition}. One such example is three-dimensional flows under solid-body rotation \cite{godeferd2015structure}, governed by the incompressible 3D rotating Navier-Stokes (3D-RNS) equations in a periodic domain,
\begin{align}
\partial_t \bm{u}+ \bm{u}\cdot \nabla \bm{u}
&= - 2\Omega \bm{\hat{z}} \times \bm{u} - \nabla p + \bm{f} + \bm{D}, ~~~~~~ \nabla\cdot \bm{u} = 0. ~~~~~ \text{3D-RNS}, \label{eq:3DRNS}
\end{align}
For such flows, a fraction of the energy injected into three-dimensional modes ($k_z\neq0$), which take the form of inertial waves, is transferred to small scales, while the remainder feeds large-scale two dimensional modes ($k_z=0$)
\cite{smith1999transfer, chen2005resonant, campagne2014direct, shaltiel2024direct} and can generate condensates \cite{seshasayanan2018condensates, clark_di_leoni_phase_2020}. 
It is a-priori unclear why condensates form in this setting; their emergence seems to require two sign-definite conserved quantities, but the flow only conserves sign-indefinite helicity and energy (like in 3D turbulence \cite{chen_physical_2003,mininni2009helicity, alexakis2017helically,galtier2003weak}).
%

Recently, a quasi-linear wave-kinetic framework was developed 
to explain how 2D condensates are maintained in rotating 3D turbulence, and to determine how energy is partitioned \cite{gome2025waves, gome2025helicity} (for earlier work using the quasi-linear framework in this context see  \cite{kolokolov_structure_2020,parfenyev2021velocity}). 
The theory in refs. \cite{gome2025waves, gome2025helicity} uses a global version of the conservation arguments presented in Sec.~4b.\ref{sec:local_2DNS}, but with the helicity of the 3D fluctuations replacing the enstrophy.
Importantly, not only do the fluctuations conserve their helicity, but rotation-dominated modes also conserve it by sign, making the invariant sign-definite. Indeed, helicity sign-mixing interactions between the waves are prohibited when mediated by the two-dimensional flow, 
as they do not satisfy the waves' resonance condition \cite{gome2025waves}. 

Here, we will build on the global results from \cite{gome2025waves, gome2025helicity} combined with ideas developed above to obtain the mean flow profile of the condensate. We will consider a frictionless case ($D=D_\nu$), with isotropic 3D forcing, in a jet geometry
(i.e. taking a box with $L_x\neq L_y$ and $L_z=L_x$).
The QL equations for the mean momentum balance \eqref{eq:momentum_balance} and the turbulent energy balance
\eqref{eq:TKE_balance} apply as is to a three-dimensional flow and are
unchanged in the presence of rotation (the Coriolis force does no work). 
Therefore, assuming scale separation as we did in Sec.~4b.\ref{sec:local_2DNS}, we expect the Reynolds stress to depend on the mean flow only through the local mean shear $U'=\partial_yU(y)$. If, in addition, the forcing is assumed to be statistically homogeneous, the remaining y-dependence in the Reynolds stress is only through $U'=\partial_yU(y)$. 
Then, from \eqref{eq:momentum_balance} we know the momentum flux must be constant:
\begin{equation}
   J_y = \langle uv\rangle\left[U'\right]-\nu U'=const,
    \label{eq:momentum_flux}
\end{equation}
which thus gives an algebraic equation for $U'$, independently of the precise functional dependence $\langle uv\rangle\left[U'\right]$~\footnote{This is a special property of frictionless jet condensates, in the presence of friction the mean flow profile will satisfy a differential equation dependent on the form $\langle uv\rangle\left[U'\right]$.}. 
We therefore obtain the solution $U'=const$ from Eq.~\eqref{eq:momentum_flux}, whether $J_y=0$ or not. 
\footnote{Note that in the presence of a constant body force acting on the mean flow, like a pressure gradient, $J_y$ depends linearly on $y$. Then, a closure $\uv \propto \epsilon/U'$ like \eqref{eq:closure} produces a logarithmic profile $U \propto \log y$ if viscosity can be neglected in the balance, akin to the log law in wall-bounded flows \cite{smits2011high}, but here in a mean-turbulent system where fluctuations are subject to an external power input of $\epsilon$ --- see \cite{nazarenko2000exact, nazarenko2000nonlinear}.} 
%
Such a profile was previously observed in simulations of rotating Rayleigh-Be\'nard convection~\cite{guervilly_jets_2017}, where turbulence is excited by convection rather than a stochastic forcing.

 \begin{figure}[t!]
    \centering
    \includegraphics[width=1\columnwidth]{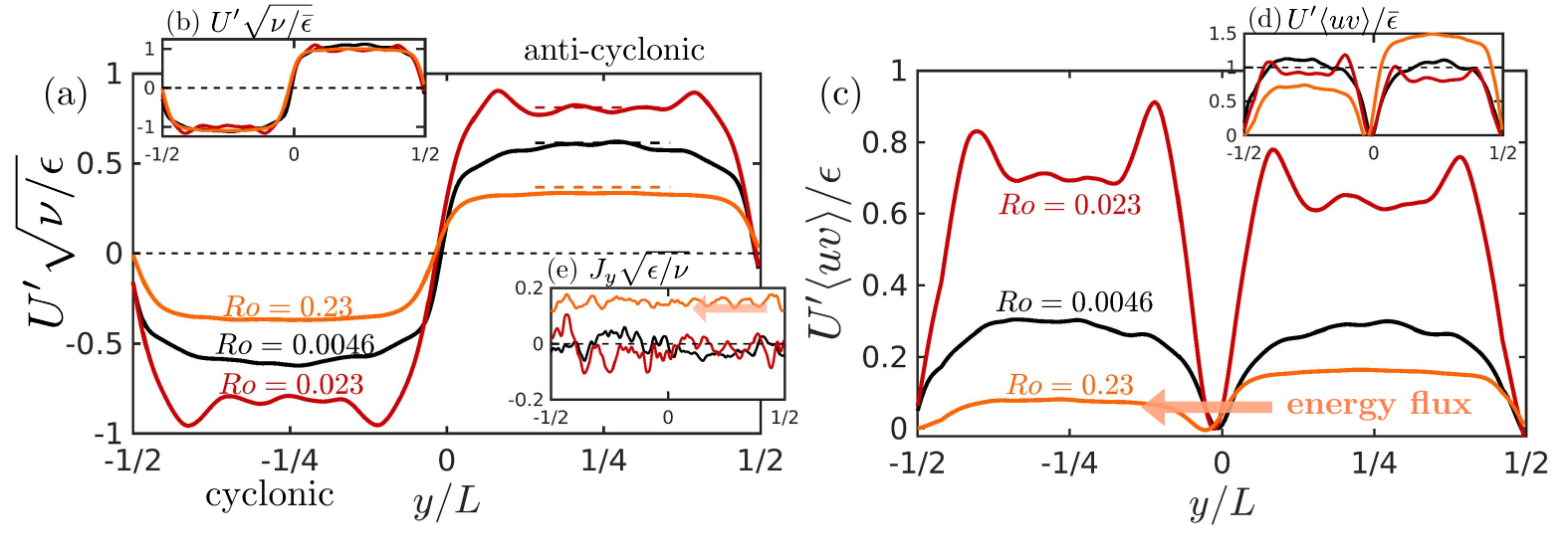}
    \caption{
        \label{fig:3DNS_nu}  Simulations of the 3D rotating Navier-Stokes equations (3D-RNS) in a jet geometry.
        (a,b) Mean-shear profile and (c,d) energy transfer to the condensate for various Rossby numbers.
        In (b) and (d), the data is rescaled with the theoretical predictions for the rms shear-rate and transfer in \cite{gome2025waves, gome2025helicity}.
        (e) Mean momentum flux $J_y$.
        A constant spatial momentum flux is observed at low rotation, producing an energy flux from anticyclonic to cyclonic regions. 
        Simulations shown: 
        \texttt{3D-RNS(A)} [Ro=0.23], \texttt{3D-RNS(B)} [Ro=0.023], \texttt{3D-RNS(C)} [Ro=0.0046].
        }
\end{figure}


What sets the amplitude of this constant mean shear rate $U'$? 
The 2DNS solution \eqref{eq:2DNS} here does not apply, as illustrated in Fig.~\ref{fig:3DNS_nu}(a): $U'(y)$ falls below $\sqrt{\epsilon/\nu}$ in the regions of constant shear. 
Instead, the amplitude of the condensate changes non-monotonically with rotation, as seen in Fig.~\ref{fig:3DNS_nu}(a) for three various Rossby numbers $Ro=\epsilon^{1/3}k_f^{2/3}/2\Omega$.
This reflects the fact that not all of the injected energy is transferred to the mean flow: at large rotation, the wave character of the 3D fluctuations reduces possible interactions with the condensate \cite{gome2025waves}, while at low rotation, some modes are rather dominated by the shear 
and transfer their energy to small scales \cite{gome2025helicity}, like in 3D turbulence.
This trend can also be seen in Fig.~\ref{fig:3DNS_nu}(c), where the local energy transfer to the condensate is shown. A theoretical estimate for the global energy transfer rate to the condensate $\bar{\epsilon}\neq \epsilon$ was obtained in ~\cite{gome2025helicity,gome2025helicity}, and when normalizing using its value, $U'/\sqrt{\bar{\epsilon}/\nu}$
indeed collapses to a value close to unity, shown in Fig.~\ref{fig:3DNS_nu}(b). 
Note that $\bar{\epsilon}$ is obtained in \cite{gome2025waves, gome2025helicity} using a global energy balance spatially-averaged over the entire box, hence corresponds to $\int (dy/L_y)\langle uv\rangle \partial_yU$, and assuming the two jets are (anti-)symmetric to leading order. One may thus expect the energy transfer to the condensate to be also locally equal to $\bar{\epsilon}$: 
$U'(y) \uv = \bar{\epsilon}$ for each jet ($\uv = \pm \sqrt{\bar{\epsilon}\nu}$
 similarly to \eqref{eq:jet_nu}).

However, that turns out to be false for the lowest rotation rate, $Ro=0.23$, where the local energy transfer to the condensate is different for the two jets, see Fig~\ref{fig:3DNS_nu}(d) (orange line). The difference stems from the existence of a non-zero momentum flux $J_y\neq0$ in this case, shown in Fig~\ref{fig:3DNS_nu}(e). Indeed, the Coriolis force breaks a symmetry of the Navier-Stokes equations: the $y\to-y$ reflection. Put another way, having the vorticity of the condensate $-\partial_yU$ aligned with the rotation direction (cyclonic jet) or anti-aligned with it (anti-cyclonic jet) can make a difference. Thus, while the momentum flux $J_y$ must be constant, it is no longer necessarily zero, leading to an asymmetry in the Reynolds stress and the associated local energy transfer, and a corresponding energy flux $J_E=  U J_y$.
Surprisingly, we observe that anti-cyclonic regions extract more energy from turbulence than cyclonic ones, and that energy is transferred from the anti-cyclonic to the cyclonic jet through the spatial flux $J_E$ ($Ro=0.23$, orange line in Fig~\ref{fig:3DNS_nu}(e)). 

We observe the approximate restoration of this symmetry at high-enough rotation (where $U'/\Omega \ll 1$ and $J_y\approx 0$), and consistently observe the same type of breaking for jets with other Reynolds numbers at low rotation. 
The origin of this symmetry breaking and its precise evolution with Ro and Re remain to be understood. 
It would be interesting to see if similar spatial energy fluxes exist in other low-rotation set-ups where asymmetry between cyclonic and anti-cyclonic structures is observed: either for vortex condensates~\cite{guervilly_large-scale_2014}, the formation of vortex crystals~\cite{clark_di_leoni_phase_2020,marchetti2025spontaneous} or homogeneous anisotropic turbulence \cite{godeferd2015structure}.

\section{SWQG: varying the Rossby deformation radius}
\label{sec:SWQG_vary_Ld}
\begin{figure}[t!]
    \centering
    \includegraphics[width=1\columnwidth]{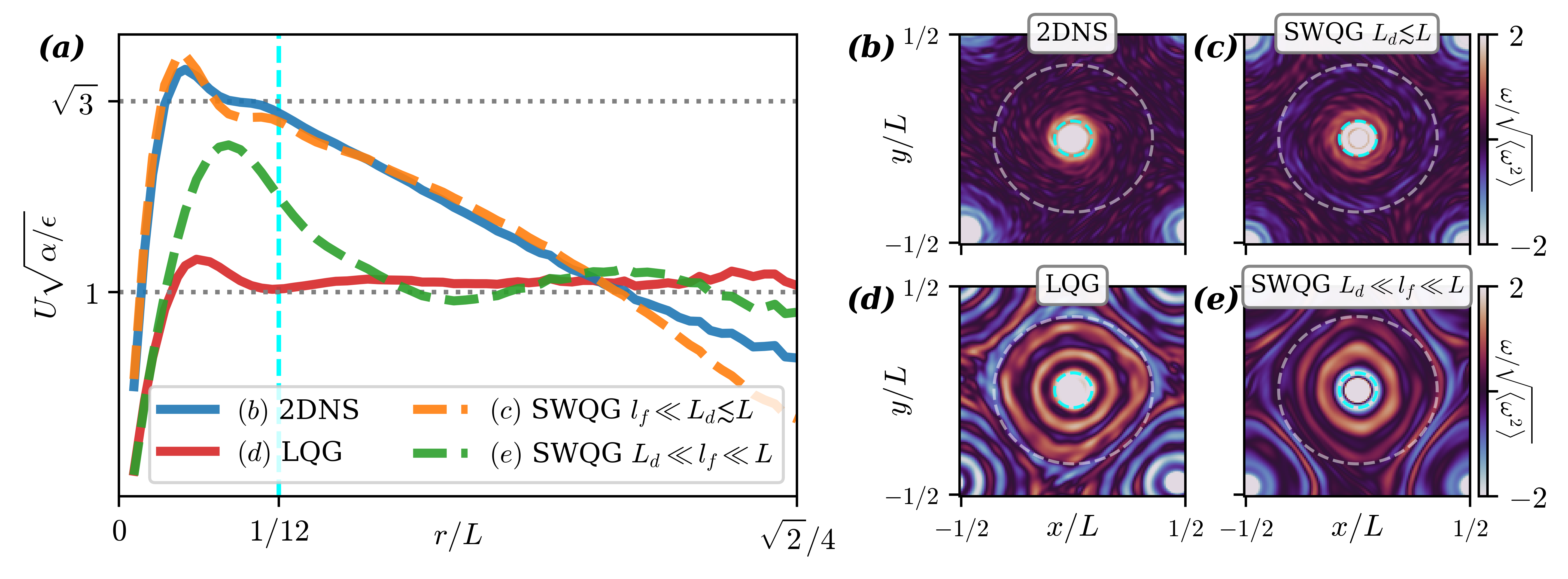}
    \caption{
        \label{fig:2DNS_LQG} 
        Condensates in (b) 2DNS, (c) SWQG at large deformation radius $L_d \sim L$, (d) 
        LQG and (e) SWQG at small deformation radius $L_d  \ll l_f \ll L$. Colors show the vorticity normalized by its rms value.
        (a) Mean vortex-condensate profile in each case.
        The white dashed circles in (b-e) indicate the regions over which the mean profiles in (a) are computed. The inner cyan circles delimit the vortex center, shown in (a) by a vertical line.
        The simulations shown are: (b) \texttt{2DNS(C)}, (c) \texttt{SWQG(A)} $[L_d/L= 0.3$, $l_f/L_d=0.2]$, (d) \texttt{LQG(D)}, and (e) \texttt{SWQG(D)} $[L_d/L= 0.04$, $l_f/L_d=3.8]$, with parameters given in Appendix~\ref{sec:simulation_details}. 
       }
\end{figure}

Having described at some length the properties of condensates in 2DNS and LQG, we now return to shallow-water QG (SWQG), Eq.~\eqref{eq:SWQG}. 
Can we learn something about condensates in the latter from condensates in the former? We certainly expect that for a large enough deformation radius $L_d = \sqrt{g_0H}/(2\Omega) \gtrsim L$ the system will behave like 2DNS. This is less clear for $L_d\ll L$, where the dynamics becomes slow when energy reaches $L_d$, leading to the formation of vortices of this typical size, see e.g.~\cite{kukharkin1995quasicrystallization, smith_turbulent_2002, boffetta2002inverse} and the recent study \cite{shi_polar_2026} including differential rotation. To the best of our knowledge, a large-scale condensate had not been previously obtained in this regime. Here, using DNS, we will show that such condensates do in fact emerge, and that they have qualitative, as well as some quantitative, similarities with condensates in 2DNS (for $L_d>l_f$) and with LQG (for $l_{\nu_p} < L_d<l_f$). This may be expected, as the conserved quantities at the forcing scale reduce to those of 2DNS for $L_d>l_f$ (with corrections being small in $(l_f/L_d)^2$) or those of LQG for $L_d<l_f$ (corrections being small in $(L_d/l_f)^2$).

We perform simulations using the Dedalus framework \cite{burns_dedalus_2020}, as detailed in Appendix.~\ref{sec:simulation_details}. In all our simulations in this section we consider a square periodic domain, $L_x=L_y$, and include friction, $D=D_\alpha$, and hyper-diffusion with $p=8$. The time scale of condensation in SWQG can be estimated as $\tau_{\mathrm{cond}} \sim (1 + (L/L_d)^2)/\alpha$, rapidly increasing as $L_d$ decreases. For this reason, and to enable comparison with LQG studies, a modest spatial resolution is chosen for the SWQG simulations. 
The main control parameter we vary is $l_f/L_d$. We observe the formation of domain-size condensates in all our simulations. 

We begin by comparing the vortex dipole condensates obtained in SWQG in two extreme limits $L_d \lesssim L$ and $L_d \ll l_f \ll L$ to condensates obtained in 2DNS and LQG, respectively. We show the mean-flow profiles in Fig.~\ref{fig:2DNS_LQG} (a) and flow visualizations in Fig.~\ref{fig:2DNS_LQG}(b-e). We also include the theoretical predictions for the vortex profile in Fig.~\ref{fig:2DNS_LQG}(a),
\begin{align}
& |V_{\phi}|=\sqrt{\frac{3\epsilon}{\alpha}}  \qquad \text{(2DNS)}, \qquad \qquad \qquad \qquad
 |V_{\phi}|=\sqrt{\frac{\epsilon}{\alpha}} \qquad \text{(LQG)}, \label{eq:vortex_profile}
\end{align}
where $|V_{\phi}|$ is the mean azimuthal velocity (see \cite{laurie_universal_2014} and section \ref{sec:profiles}).
We find that the profile of the SWQG condensate for $L_d \lesssim L$ (dashed orange in Fig.~\ref{fig:2DNS_LQG}(a)) agrees quantitatively with that of the 2DNS vortex (blue line in Fig.~\ref{fig:2DNS_LQG}(a)).
Note that the predicted 2DNS profile \eqref{eq:vortex_profile} is here expected to hold in a small fraction of the domain (outside which the local QL condition breaks, see Sec. 4~\ref{sec:assumptions}),
delimited by the teal dashed line in Fig.~\ref{fig:2DNS_LQG}(a). This small region of consistency is also observed for SWQG.
In contrast, SWQG with $L_d \ll l_f \ll L$ exhibits an LQG-like condensate (Fig.~\ref{fig:2DNS_LQG}(e)), the profile following the LQG prediction in \eqref{eq:vortex_profile} instead. The core of the vortex however resembles more closely that of 2DNS, though the difference from LQG could also be due to the different  values of $l_f$ and $\nu_p$ in the two. We also note the presence of 
annular modulations of the vortex condensate
in both SWQG examples, characteristic of LQG. They are more pronounced in the LQG-like case, panel (e), which also exhibits slight symmetry breaking, typical of LQG.

\begin{figure}[t!]
    \centering    \includegraphics[width=1.0\columnwidth]{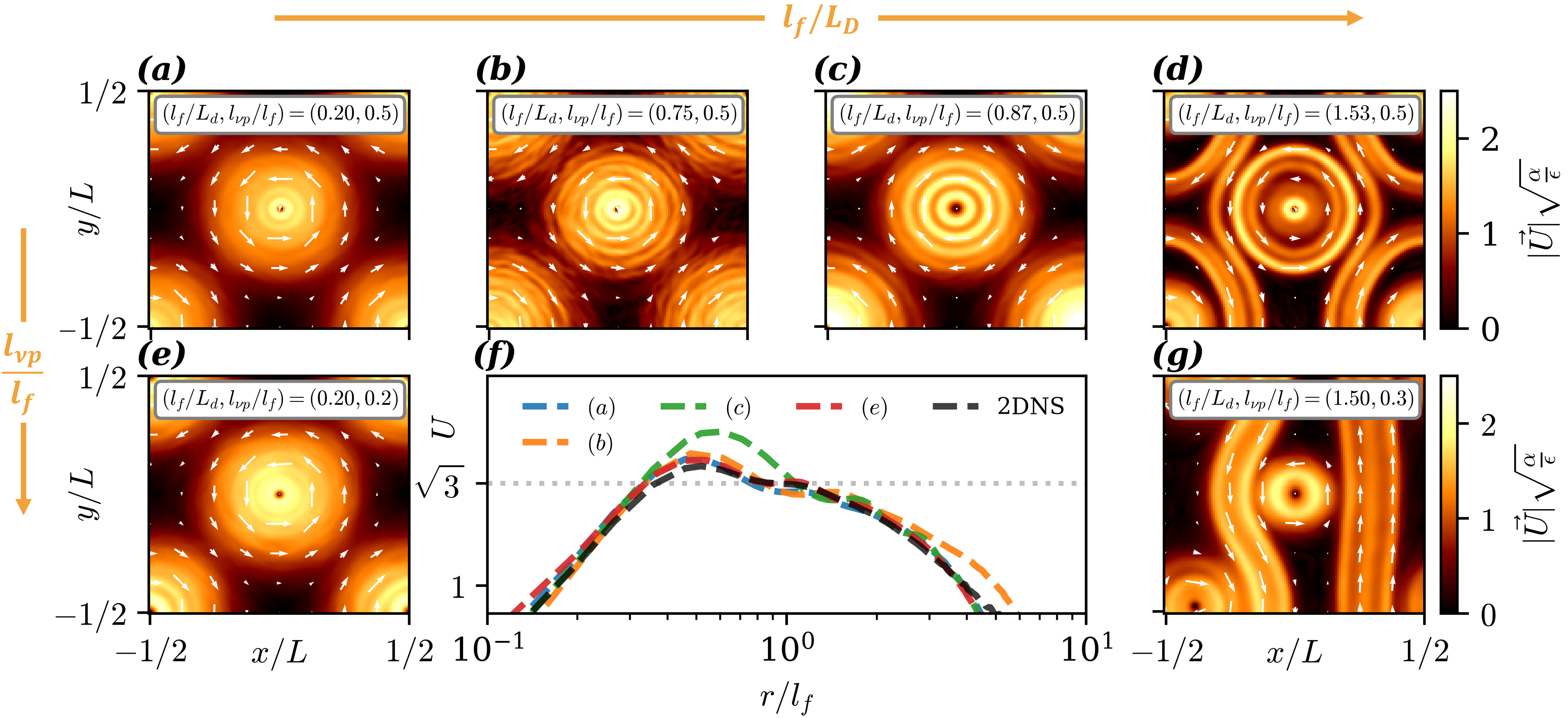}
    \caption{
    Condensates in SWQG with various Rossby deformation radii ($l_f/L_d$ increasing from left to right), and ranges for the forward cascade ($l_{\nu_{p}}/l_f$ decreasing from top to bottom).
    Colors and arrows in (a-e,g) show, respectively, the velocity magnitude and its direction. 
    (f) Mean vortex profiles for $l_f/L_d < 1$ (a–c,e). For reference, the 2DNS vortex profile (black) and the prediction from \eqref{eq:vortex_profile} (gray) are shown.  
        The simulations shown are: (a) \texttt{SWQG(A)}, (b) \texttt{SWQG(F)}  (c) \texttt{SWQG(I)}, (d) \texttt{SWQG(B)}, (e) \texttt{SWQG(L)}, (g) \texttt{SWQG(K)} and \texttt{2DNS(C)} with parameters given in Appendix~\ref{sec:simulation_details}. 
    }
    \label{fig:SWQG_lf+Ld+l_nu}
\end{figure}
Having discussed the two 
limiting SWQG cases,
we turn to explore the changes SWQG condensates undergo with increased $l_f/L_d$, see Fig.~\ref{fig:SWQG_lf+Ld+l_nu}. Simulations of SWQG are evolved from zero initial conditions~\footnote{A regime of bistablity for condensate configurations was found in LQG, here we did not explore this issue.}, and the resulting condensates are shown for different values of $l_f/L_d$, with $l_{\nu_p}/l_f = 0.5$ kept constant in Fig.~\ref{fig:SWQG_lf+Ld+l_nu}(a)-(d). We observe that for $0.2 < l_f/L_d < 0.87$ the condensate resembles a 2DNS-like vortex (panels a-c), agreeing with the profile given by \eqref{eq:vortex_profile}. Note that even though some of these simulations have small deformation radii $L_d/L\sim0.1$, they are still very similar to 2DNS.
The only qualitative difference from 2DNS is the presence of the LQG-characteristic 
annular modulations within the vortex, becoming more distinct as $l_f/L_d\to 1$. 
It is only when $l_f/L_d$ increases above one that the profile of the condensate begins to resemble the LQG condensate (panel d). 

As discussed in Sec.4b \ref{sec:KE_LQG}, in LQG the direct cascade is suppressed in regions where the condensate is strong, and occurs in complementary regions where the condensate is absent. This partitioning, combined with the requirement that kinetic and potential energy injection and dissipation be balanced, places a constraint on the area the condensate can occupy. This constraint becomes increasingly limiting as the extent of the direct inertial range $l_{\nu_p}/l_f$ (where $l_{\nu_p}\ll l_f$ is the hyperviscous dissipative scale) decreases, eventually leading to spontaneous symmetry breaking and the formation of jets \cite{svirsky_out--equilibrium_2025}. This does not occur in 2DNS. 
We now test if the same phenomenology is observed in SWQG. Decreasing $l_{\nu_p}/l_f$ for simulations with $L_d<l_f$ indeed induces a transition to a jet condensate, similar to the one observed in LQG in a square domain, see Figs.~\ref{fig:SWQG_lf+Ld+l_nu}(d) and (g). In contrast, lowering $l_{\nu_p}/l_f$ produces no noticeable effect in the 2DNS-like cases with $l_f/L_d <1$, see Figs.~\ref{fig:SWQG_lf+Ld+l_nu}(a) and (e).

\begin{figure}[t!]
    \centering
    \includegraphics[width=1\columnwidth]{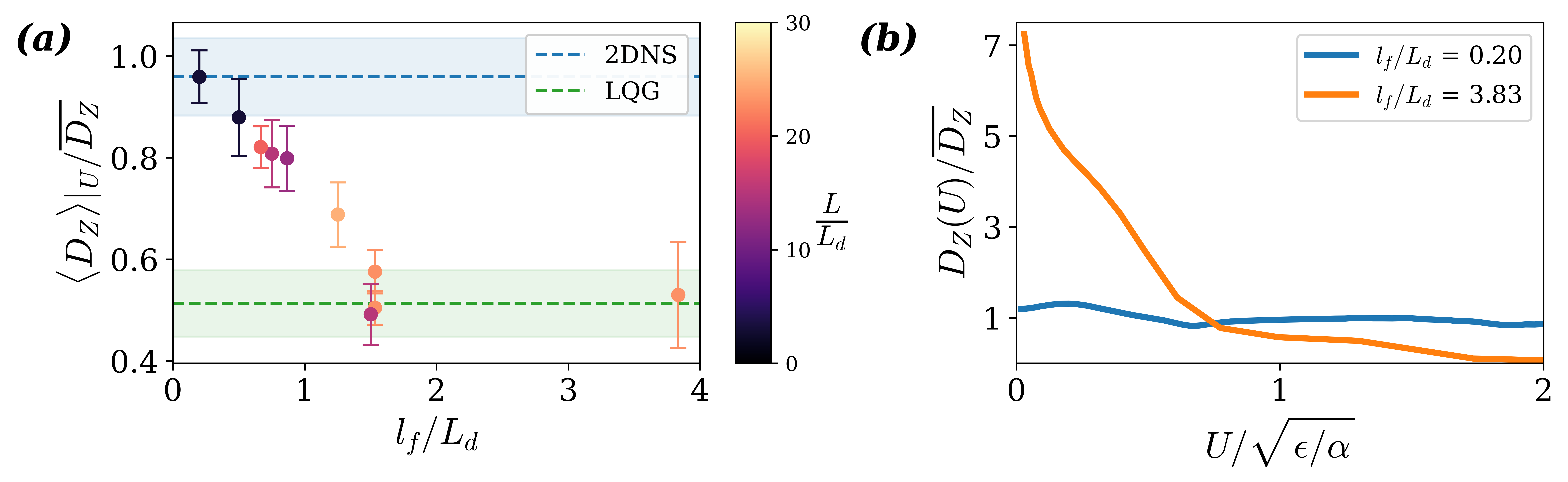}
    \caption{
      (a) Average local viscous dissipation of $Z_q$ in SWQG condensates in steady state, conditioned on regions with strong mean flow ($U>0.8\sqrt{\epsilon/\alpha}$). Shown are values normalized by the mean dissipation $\overline{D_Z}$ (unconditioned). Points correspond to the mean value, with error bars representing one standard deviation across steady-state snapshots. 
      The horizontal lines show the reference mean values computed for 2DNS and LQG, with shaded margins indicating one standard deviation of the mean. Colors indicate the value of the inverse deformation scale $1/L_d$. (b) IIllustration of the dependence of the local viscous dissipation on the condensate amplitude. Shown is the average local viscous dissipation conditioned on specific values of the local flow amplitude, $D_Z(U) \equiv \av{D_Z}_{U \leq \abs{\bm{u}} < U+dU}$, for the two most extreme values of $l_f/L_d$ shown in panel (a).
      Simulations used (ordered by increasing $l_f/L_d$): 
      \texttt{SWQG (A),(C), (H), (F), (I), (G), (K), (E), (B){} and \texttt{(D)}.  
    }}
    \label{fig:swqg_DZ}
\end{figure}

The latter results hint that the influence of the condensate on the direct cascade in SWQG is similar to that in 2DNS and LQG in the respective regimes. To quantitatively assess this, we measure the local (hyper-)viscous dissipation of the squared potential vorticity, $Z_q=\int q^2/2\dd^2x$, for the fluctuations $D_Z(\bm{x})$ (as $l_{\nu_p}<L_d$ the dominant contribution to the dissipation of $Z_q$ for the fluctuations comes from small scales $\sim l_{\nu_p}$). 
We then compute the conditional average $\av{D_Z}|({U>\theta\sqrt{\epsilon/\alpha}})$ 
that is, the dissipation in regions where the velocity magnitude exceeds a given threshold. This conditional average is computed for each snapshot in steady state, and normalized by the space averaged dissipation $\bar{D}_Z$. We choose the threshold $\theta = 0.8$, so that regions with velocities comparable to the condensate magnitude are included, while accounting for fluctuations about the mean. The dependence of the dissipation on $U$ is demonstrated in Fig.~\ref{fig:swqg_DZ}(b) for two limiting cases, while $\av{D_Z}|({U>0.8\sqrt{\epsilon/\alpha}})/\bar{D}_Z$ as a function of $l_f/L_d$ is shown in Fig.~\ref{fig:swqg_DZ}(a).

 For $l_f/L_d=0.2$ the local dissipation of $Z_q$ essentially equals $1$, as it does in 2DNS, dropping somewhat for larger $l_f/L_d$, but still remaining $\sim 1$ (Fig.~\ref{fig:swqg_DZ}(a)),
indicating that local dissipation of $Z_q$ almost balances local injection\footnote{In Fig.~\ref{fig:swqg_DZ} we normalize by the mean dissipation $\overline{D}_Z$. The total dissipation of $Z_q$ also includes a small large-scale contribution from $D_\alpha$; however, $\overline{D}_\alpha/\overline{D}_Z \lesssim 0.03$ for all cases considered, so that $\overline{D}_Z= \eta$ to a good approximation.}, and that there is little suppression of the direct cascade in the condensate region for $l_f/L_d<1$ . The dissipation decreases more significantly as $l_f/L_d$ increases above one. It eventually saturates at values of order $\sim 0.5$, consistent with those observed in LQG, implying significant suppression of small-scale generation in condensate regions. Note that this effect is independent of $L_d/L$, confirming that $l_f/L_d$ is indeed the relevant control parameter. 

The one distinguishing feature of SWQG with $l_f/L_d<1$ from 2DNS is the flow undulations, also observed in LQG. They are visible in flow snapshots but not in averages, and correspond to modes for which the advection by the condensate identically vanishes (radially symmetric modes for vortices, $k_x=0$ modes for jets). Such zero modes are directly forced in our simulations, and can interact only with other fluctuations but not with the condensate.
 In LQG, however, fluctuations that interact with the condensate essentially remain at the scale $l_f$, which restricts the possible triads that the zero modes can participate in. Thus, zero modes also mostly remain at the forcing scale, with their energy injection and dissipation balanced there. This can explain their large amplitude~\footnote{The amplitude can be estimated from these considerations. We have verified the resulting amplitude agrees with measurements in LQG simulation (A-C), with different $\alpha$ values but same $k_f$ (not shown)}, and is consistent with our observations that $l_f$ is their characteristic scale (see \cite{SM} for future details). This is also consistent with such modes becoming apparent in SWQG only once small-scale dissipation becomes suppressed in condensate regions, as $L_d$ approaches $l_f$: compare \texttt{SWQG(A)} (Figs.~\ref{fig:SWQG_lf+Ld+l_nu}(a)) and \texttt{SWQG(F)} Figs.~\ref{fig:SWQG_lf+Ld+l_nu}(b) with their dissipation values in Fig.~\ref{fig:swqg_DZ}(a).




\section{Conclusion}
\label{sec:conclusion}
Across various examples of self-organized, inhomogeneous turbulence, an underlying universal structure emerges. The interplay between the two sign-definite invariants of the flow leads to anti-diffusive behavior and shapes the resultant large-scale flow. In the presence of extended inertial ranges, small scales experience universal dynamics, and are dominated by the large-scale flow. A perturbative, self-consistent approach then becomes possible, enabling closure and the successful prediction of the mean-flow profile in several examples. A dichotomy between PT (Parity + Time reversal) invariant and PT breaking second-order moments is revealed: the former are determined at large scales, sensitive to the full mean-flow profile, while the latter are determined at small scales, feeling only the local shear, and crucially depend on PT breaking terms (such as forcing). It would be interesting to see if some of these ideas could find applications for three-dimensional turbulence in the presence of mean flows, e.g. for wall-bounded turbulence which is self-sustained rather than externally driven~\cite{farrell2012dynamics, thomas2015minimal,hwang_attached_2020}. 

So far, we have emphasized the successes of the theoretical approach. It is also worthwhile pointing out several challenges. First, the theory does not predict the global structure of the mean flow; it assumes condensation at the largest scale, though there are cases where scale selection is known to occur~\cite{vallis_generation_1993,clark_di_leoni_phase_2020}. 
Typically, the theory also does not apply in the entire domain: either because the assumed symmetries are broken (e.g. hyperbolic regions between vortices), or the QL assumption is broken (in direct cascade regions in LQG, regions of weak shear in 2DNS, or eddy-eddy interactions being negligible only at scales larger than forcing scale, e.g. as in~\cite{galperin_anisotropic_2006}).
Furthermore, the mean flow profile obtained theoretically could have instabilities, which can generate additional fluctuations on top of those directly excited by the forcing, not captured within the theory, see e.g. in the context of jets on the beta plane~\cite{constantinou_emergence_2014,galperin_zonal_2019,jackman_parameterisation_2023} and possibly in 2DNS~\cite{frishman_jets_2017}. 
Progress on these issues requires a better understanding of the interplay between mean-flow dominated dynamics and wave/eddy dominated-dynamics, coexisting in different spectral or spatial regions, combined with more dynamical QL theories~\cite{bouchet_kinetic_2013,  michel2019multiple, chini_exploiting_2022}, which is an exciting direction for the future.


\dataccess{All data used to perform this study and the code to generate the presented figures will be made available upon request.}


\competing{The authors declare that they have no competing interests.}

\funding{A.F is supported by BSF grant No. 2022107 and ISF grant No. 486/23.}

\ack{A.F. thanks Tsuf Lichtman for useful discussions.}



\vskip2pc


\bibliography{jet_bib}

\appendix
\section{Simulation details}
\label{sec:simulation_details}

For 2DNS, LQG, and shallow-water QG (SWQG), we perform direct numerical simulations (DNS) using the Dedalus (v3) framework \cite{burns_dedalus_2020}. The equations are solved with a pseudo-spectral method using the 3/2 dealiasing rule, and time integration is carried out with a third-order, four-stage DIRK/ERK scheme. The parameters used are given in Table~\ref{tab:sim-parameters}.

The simulations use a variable time step constrained by the CFL condition, with the maximum Courant number set to $C_{\text{max}} = 0.5$. The flow is driven by a white-in-time forcing that is localized in Fourier space at wavenumber $k_f = 2\pi/l_f$. The forcing is applied within an annulus $k_f - 1 \leq k \geq k_f +1$, a constant amplitude $A_f$ and a random phase. 
The amplitude is chosen such that the energy injection rate $\epsilon$ is comparable across simulations of 2DNS and SWQG, yielding  $\delta = \alpha L^{2/3}/\epsilon^{1/3} \sim 10^{-3}$ for all simulations with friction. In frictional 2DNS, this corresponds to $R_u/L = \delta^{-1/2}k_f^{-2/3} \approx 2$, where $R_u$ denotes the radius at which the nonlinear interaction rate becomes comparable to the mean-flow shear rate \cite{frishman_turbulence_2018}. 
%
In all simulations, we chose $L_y = L = 2\pi$ and $L_x = L_y (N_x/N_y)$.
Periodic boundary conditions are imposed in both spatial dimensions.

Three dissipation mechanisms are included, each with its own coefficient: linear drag $\alpha$, regular viscosity $\nu$, and hyper-viscosity $\nu_p$. All simulations start from rest and are integrated until they reach a statistically steady state, identified by the saturation of the inverse-cascading quadratic invariant $E$. The condition for steady state is $|( \tau_L / E )  \partial E/\partial t| < 0.1$, where $\tau_L$ is the eddy turnover time at the box scale. Once steady state is reached, statistics are collected for at least $T \ge 200 \tau_L$.

For 3DRNS, we use the pseudo-spectral code GHOST \cite{mininni2011hybrid} to simulate the primitive equations \eqref{eq:3DRNS} in a 3D domain $L_x=L_z= L_y/2 $ ($N_z=N_x$), and with 3D isotropic forcing. The CFL number is lowered (taken to $0.2$) in order to resolve the fast inertial waves generated by the Coriolis force. 

\begin{table}[t]
\caption{\label{tab:sim-parameters} Parameters of the simulations conducted. The prefixes 2DNS, LQG, SWQG and 3DRNS identify the equation integrated in each run. The table lists the following parameters: forcing scale $k_f = 2\pi/l_f$, deformation-radius scale $k_d = 2\pi/L_d$, rotation rate $\Omega$, linear drag coefficient $\alpha$, viscosity coefficient $\nu$, hyper-viscosity coefficient $\nu_p$ with exponent $p$, spatial resolution $N_x, N_y$, forcing amplitude $A_f$, and the resulting energy injection rate $\epsilon$.
}
\begin{tabular}{lccccccccccc}
\textrm{}&
\textrm{ $k_f$ }&
\textrm{ $k_d$ }&
\textrm{ $\Omega$ }&
\textrm{ $\alpha$ }&
\textrm{ $\nu$ }&
\textrm{ $\nu_p$ }&
\textrm{ $p$ }&
\textrm{ $N_x$ }&
\textrm{ $N_y$}&
\textrm{ $A_f$}&
\textrm{ $\epsilon$ }\\
\hline
2DNS(A)& 100 &  - &  - & $1\cdot10^{-3}$ &   -   & $1\cdot10^{-33}$ & 8 & 512 & 512 &  12 &               0.17 \\ 
2DNS(B)& 50 &  - &  - & $1\cdot10^{-3}$ &   -   & $1\cdot10^{-34}$ & 8 & 512 & 512 & 6.5 &               0.11 \\ 
2DNS(C)& 15 &  - &  - & $1\cdot10^{-3}$ &   -   & $6\cdot10^{-24}$ & 8 & 128 & 128 &   3 &               0.12 \\ 
2DNS(D)&  50 &  - &  - &            -  &$10^{-3}$& $6\cdot10^{-24}$ & 8 & 128 & 256 &  30 &               10.8 \\ 
LQG(A)&  13 &  - &  - & $2\cdot10^{-3}$ &   -   & $7\cdot10^{-19}$ & 7 &  64 & 128 & 0.2 &  $2.44\cdot 10^{-4}$ \\ 
LQG(B)&  13 &  - &  - & $1\cdot10^{-3}$ &   -   & $7\cdot10^{-19}$ & 7 &  64 & 128 & 0.2 &  $2.42\cdot 10^{-4}$ \\ 
LQG(C)&  13 &  - &  - & $5\cdot10^{-4}$ &   -   & $7\cdot10^{-19}$ & 7 &  64 & 128 & 0.2 &  $2.42\cdot 10^{-4}$ \\ 
LQG(D)&  10 &  - &  - & $1\cdot10^{-3}$ &   -   & $1\cdot10^{-17}$ & 7 & 128 & 128 & 0.2 &  $4.55\cdot 10^{-4}$ \\ 
SWQG(A)&  15 &  3 &  - & $1\cdot10^{-3}$ &   -   & $6\cdot10^{-24}$ & 8 & 128 & 128 &  3  &               0.12 \\ 
SWQG(B)&  15 & 23 &  - & $1\cdot10^{-3}$ &   -   & $6\cdot10^{-24}$ & 8 & 128 & 128 &  3  &               0.12 \\ 
SWQG(C)&   6 &  3 &  - & $1\cdot10^{-3}$ &   -   & $6\cdot10^{-24}$ & 8 & 128 & 128 &  3  &               0.75 \\ 
SWQG(D)&   6 & 23 &  - & $1\cdot10^{-3}$ &   -   & $6\cdot10^{-24}$ & 8 & 128 & 128 &  3  &               0.75 \\ 
SWQG(E)&  15 & 23 &  - & $1\cdot10^{-2}$ &   -   & $6\cdot10^{-24}$ & 8 & 128 & 128 &  3  &               0.12 \\ 
SWQG(F)&  20 & 15 &  - & $5\cdot10^{-2}$ &   -   & $6\cdot10^{-26}$ & 8 & 128 & 128 &  4  &               0.16 \\ 
SWQG(G)&  20 & 25 &  - & $5\cdot10^{-2}$ &   -   & $6\cdot10^{-26}$ & 8 & 128 & 128 &  4  &               0.16 \\ 
SWQG(H)&  30 & 20 &  - & $5\cdot10^{-2}$ &   -   & $6\cdot10^{-26}$ & 8 & 128 & 128 &  3  &               0.03 \\ 
SWQG(I)&  15 & 13 &  - & $1\cdot10^{-3}$ &   -   & $6\cdot10^{-26}$ & 8 & 128 & 128 &  3  &               0.12 \\ 
SWQG(J)&  15 &0.1 &  - & $1\cdot10^{-3}$ &   -   & $6\cdot10^{-26}$ & 8 & 128 & 128 &  3  &               0.12 \\ 
SWQG(K)&  10 & 15 &  - & $5\cdot10^{-2}$ &   -   & $6\cdot10^{-26}$ & 8 & 128 & 128 &  3  &               0.27 \\ 
SWQG(L)&  3  & 15 &  - & $1\cdot10^{-3}$ &   -   & $6\cdot10^{-30}$ & 8 & 256 & 256 &  3  &               0.12 \\ 
SWQG(M)&  23 & 15 &  - & $1\cdot10^{-3}$ &   -   & $6\cdot10^{-30}$ & 8 & 256 & 256 &  3  &               0.12 \\ 
3DRNS(A)&  10 &  -  & 10  & -            & $10^{-3}$ & $6\cdot10^{-29}$ & 8 & 128 & 256 &  30  &      1 \\
3DRNS(B)&  10 &  -  & 100 & -            & $10^{-3}$ & $6\cdot10^{-29}$ & 8 & 128 & 256 &  30  &      1 \\
3DRNS(C)&  10 &  -  & 500 & -            & $10^{-3}$ & $6\cdot10^{-29}$ & 8 & 128 & 256 &  30  &      1 
\end{tabular}
\end{table}

\newpage
\section*{Supplementary material
}

\subsection{Varying the strength of bottom drag at the forcing scale}
\label{apx:varying_drag_strength_at_lf}

\begin{figure}
    \centering
    \includegraphics[width=1.0\columnwidth]{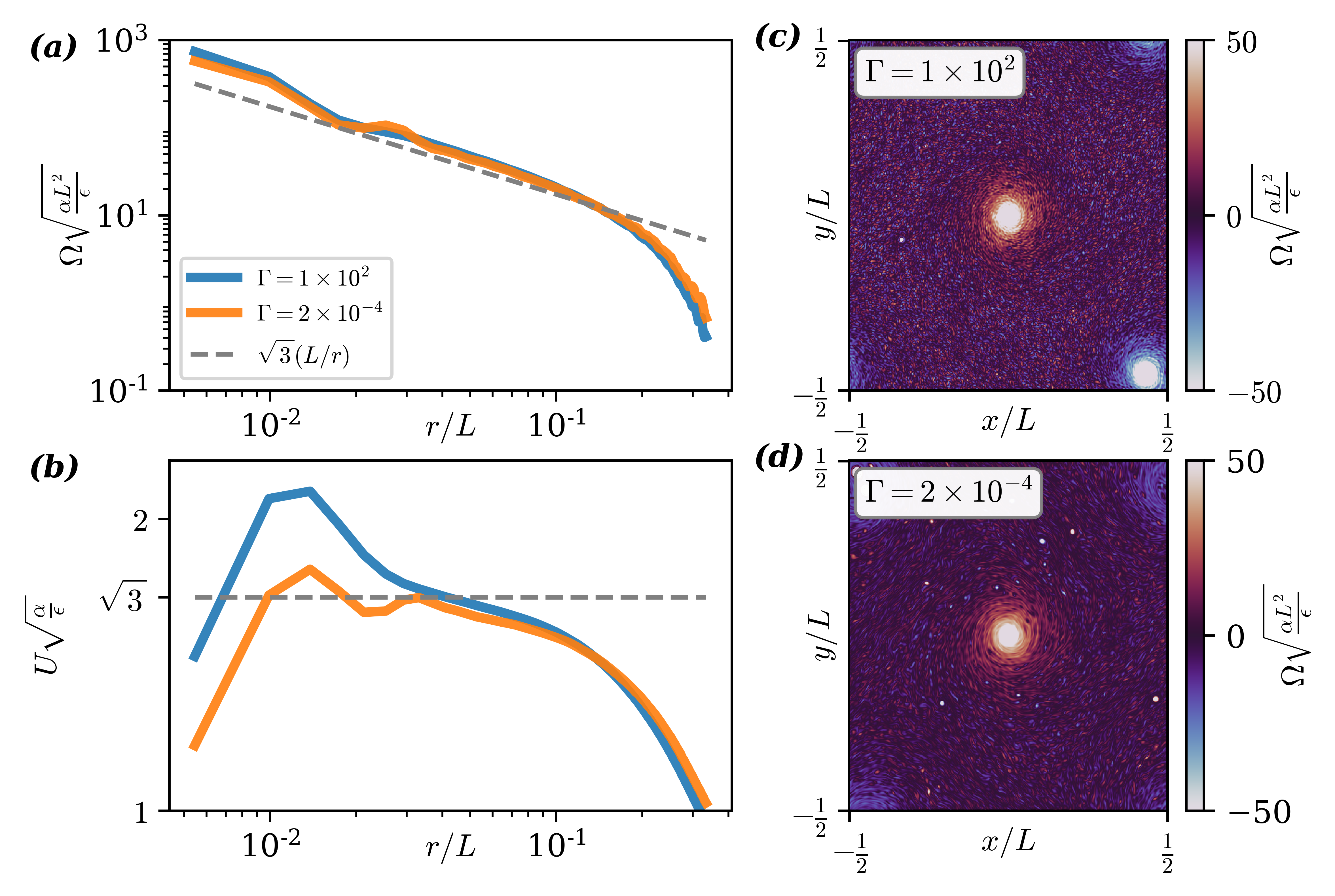}
    \caption{
        \label{fig:different_gamma} 
        Simulation results for 2DNS with linear friction ($D = D_\alpha$) for different values of 
        $\Gamma = \nu_p k_f^{2p}/\alpha$ -- the ratio of viscous dissipation to bottom drag at the forcing scale. 
        (a) Mean vorticity profile $\Omega$. 
        (b) Mean azimuthal velocity profile $U$. 
        (c,d) Snapshots of the vorticity field in the regimes where viscous dissipation dominates at the forcing scale (c) and where bottom friction dominates (d). In (a-b), dashed gray lines correspond to the QL prediction for the condensate (Eq.~(8.1) in the main text).
        Simulations shown: (c) \texttt{2DNS(A)} $[\Gamma = 10^{2}]$; (d) \texttt{2DNS(B)} $[\Gamma = 2\cdot 10^{-4}]$, with parametrs given in Appendix.~A in the main text.
        }
\end{figure}

Here we expand the discussion regarding the validity of the closure in 2DNS when friction dominates over viscosity at the forcing scale. In the main text, the closure 
\begin{equation}
    \label{eq:apx_closure}
     U' (y) \langle uv\rangle (y)=\epsilon
\end{equation}
is derived for 2DNS with viscosity as the main dissipation mechanism $D=D_\nu$. If instead, bottom drag is the main dissipation mechanism (in particular $D=D_\alpha$), while viscous dissipation is negligible at the forcing scale ($\Gamma = \nu_p k_f^{2p}/\alpha \ll 1$), a more subtle discussion is required, ~\cite{Srinivasan2012,srinivasan_reynolds_2014,kolokolov_structure_2016,frishman_culmination_2017,jackman_parameterisation_2023}. Linear friction corresponds to a temporal cutoff $\propto 1/\alpha$, instead of a wavenumber cutoff. This is a crucial distinction when considering the dynamics of the fluctuations: modes which start with small $k_x$ and are initially aligned against the shear ($k_yk_x U'<0$), are first sheared towards smaller $k_y$. During this stage, the corresponding energy flux $\Pi^E$ increases to large values (since $k^2$ decreases), corresponding to the extraction of energy from the condensate by large-scale fluctuations (the Orr-mechanism, see, e.g., \cite{farrell1993optimal, bouchet_kinetic_2013}). 

Those modes are eventually sheared over, but only after a time of order $t\sim k_f/(k_x U')$. The requirement $U'/\alpha \gg L/l_f$ is sufficient to guarantee that energy is able to reach small enough scales, $k_d\gg k_f$, so that all modes $k_x$ have a decaying energy flux (except for the isolated mode $k_x=0$), see also the discussion in~\cite{frishman_culmination_2017,jackman_parameterisation_2023}. This condition is equivalent to the requirement $\delta^{1/2}(L/l_f)\ll1$, which is only slightly more stringent than the QL condition derived above. 

Correspondingly, our DNS results, shown in Fig.~\ref{fig:different_gamma}S indicate that when $D=D_\alpha$, whether friction or hyper-viscosity dominate around the forcing scale, does not alter the condensate, and the result obtained for the closure \eqref{eq:apx_closure} seems to hold.

Note that if no IR cutoff is assumed for $k_x$, while the friction is kept finite, then $ k_x U'/\alpha < k_f$ for a fraction of the excited modes which will never be sheared over (corresponding to modes with $\phi=\tan^{-1}(k_y/k_x)\sim \pm \pi/2$, depending on the sign of $U'$). They will thus extract energy from the condensate rather than transferring their energy to it.

It turns out that for an isotropic forcing, and for any arbitrarily small $\alpha$, the total Reynolds stress $\langle uv\rangle=0$: the correspondingly small neighborhood of modes around $k_x=0$ extract an amount of energy from the condensate which exactly balances the fraction of energy transferred to the condensate ~\cite{srinivasan_reynolds_2014,kolokolov_structure_2016}. In this scenario, energy is still transferred to large scales, but to large-scale \textit{fluctuations}, the condensate only serving as a mediator: energy extracted by the condensate from most of the excited modes is transferred by it to a vanishingly small fraction of large-scale fluctuating modes. Similarly to the discussion in~\cite{kraichnan_eddy_1976}, such cancellations of the energy flux when integrated over $k_x$ can be traced to a Lagrangian conservation law $\int \frac{d\phi}{2\pi} (k_f^2/k^{2}(t))=1$~\cite{frishman_statistical_2015}, here written for the advection of the wavenumber along characteristics and taking $k_y(t)\sim k_y(1/\alpha)$, starting from an isotropic distribution on a circle of radius $k_f$.  

\subsection{Origin of the sub-jet structure}

\begin{figure}[t!]
    \centering
    \includegraphics[width=1\columnwidth]{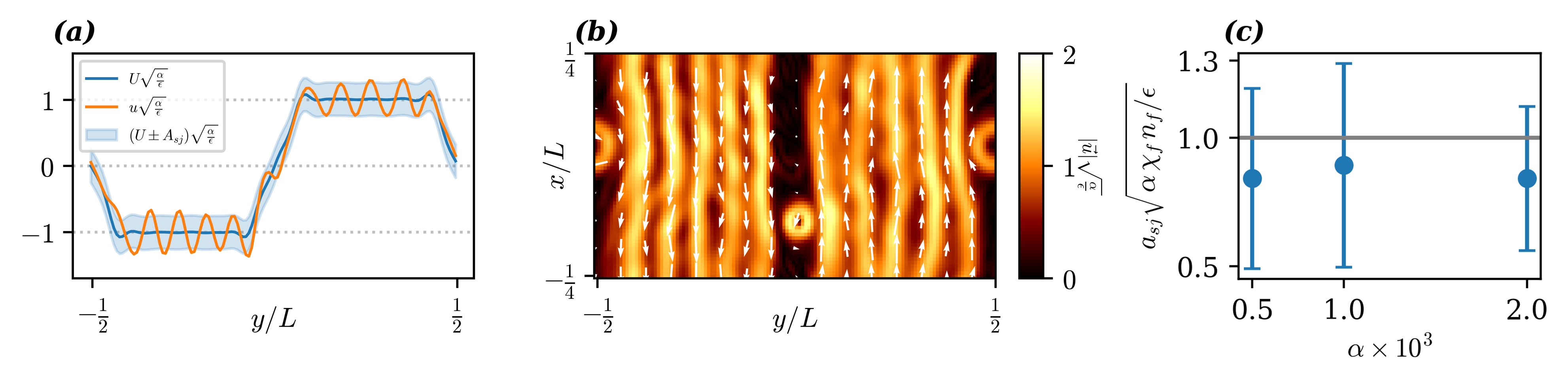}
    \caption{
        Sub-jet structure in an LQG condensate in a rectangular domain. 
        (a) Time-averaged horizontal velocity profile $U$, with the shaded region indicating the predicted sub-jet amplitude $A_{sj}$, and a snapshot of the horizontal profile $u$. 
        (b) Snapshot of the velocity field, with color representing velocity magnitude and white arrows indicating direction.
        The $u$ profile in (a) corresponds to the $x$-average of the snapshot in (b). (c) The mean square amplitude of the velocity fluctuations $u - U$ for different values of the drag coefficient $\alpha$, normalized by the prediction \eqref{eq:square_amp_prediction}. 
        Simulation shown:(a-b) \texttt{LQG(B)}, (c) \texttt{LQG(A)}$[\alpha = 2 \cdot 10^{-3}]$,\texttt{LQG(B)}$[\alpha = 10^{-3}]$,\texttt{LQG(C)} $[\alpha = 0.5\cdot 10^{-3}]$  with parameters given in Appendix.~A of the main text.
    }
    \label{fig:LQG_subjets}
\end{figure}

Here we address a common feature of LQG-like condensates: the presence of  modulations of the large-scale flow, visible in Figs.~5 and Fig.~6 in the main text. These undulations also appear in LQG and are a robust feature. Their wavelength is comparable to $l_f$, suggesting that they are confined to the forcing scale. To explain their origin, we turn to LQG and focus on the jet condensate, where the simpler geometry allows for a cleaner analysis.

In Fig.~\ref{fig:LQG_subjets}, an example of an LQG jet condensate in a rectangular domain is shown. The modulation, or sub-jets, are clearly visible in a snapshot of the flow.

Considering LQG (Eq.~2.3, in main text) with friction in $k$-space with $\bm{k} = (k_x,k_y)$. We assume jet geometry with $\bm{\nabla} \Psi  = \sqrt{\epsilon/\alpha} \hat{x}$, and employing a quasi-linear approximation, we write the leading-order balance for modes homogeneous in $x$ (with $k_x=0$), denoted $\hat\psi_{(0,k_y)} \equiv \hat\psi(\bm{k}=(0,k_y))$. By symmetry of the condensate, the leading-order nonlinear interactions between the condensate and the $\hat\psi_{(0,k_y)}$ modes vanish. Additionally,  since modes with scales smaller than the forcing cannot be generated where the condensate is strong, and non-linear interactions of the $k_x=0$ modes with other modes on the forcing ring are rather restricted, the dynamics of these modes essentially becomes
\begin{equation}
    \partial_t \hat\psi_{(0,k_y)} = \hat{g}_{(0,k_y)}(t) - \alpha k_y^2 \hat\psi_{(0,k_y)},
\end{equation}
where $\hat{g}_{\bm{k}}$ is white-in-time forcing centered at wavenumber $k_f$, with $\av{\hat g_{\bm{k}}(t) \hat g_{\bm{k'}}^*(t')} = 2 \epsilon \delta_{\bm{k},\bm{k'}} \hat \chi_{\bm{k}}\delta(t - t')$, and $\hat \chi_{\bm{k}}$ is then the density of forced modes. In steady state, this yields $\av{|\hat \psi_{(0,k_y)}|^2} = \epsilon \hat \chi_{(0,k_y)}/(k_y^2 \alpha)$ and hence $\av{ |\hat u_{(0,k_y)}|^2 } = \epsilon  \hat \chi_{(0,k_y)}/\alpha$, which vanishes for all $k_y$ that are not directly forced.

In simulations, the forcing is applied not to a single mode but to a narrow annulus around $k_f$ (see Appendix.~A in main text). We therefore expect the squared amplitude of the sub-jets to scale as
\begin{equation}
    A_{sj}^2 = \frac{\epsilon}{\alpha} \sum_{k_y} \hat\chi_{(0,k_y)},
\end{equation}
where we have summed over the contributions of all forced modes. In Fig.~\ref{fig:LQG_subjets}, simulation results are shown demonstrating that the amplitude of the fluctuations is well predicted by
\begin{equation}
    \label{eq:square_amp_prediction}
    A_{sj}^2 = \frac{\epsilon}{\alpha}  \chi_{(0,k_f)} n_f,
\end{equation}
where $n_f=3$ is the number of forced modes and  $\chi_{(0,k_f)} \approx 1.5\cdot10^{-2}$. 

This prediction can be tested quantitatively in simulations with different values of $\alpha$, as shown in Fig.~\ref{fig:LQG_subjets}(c), where we measure the amplitude of the fluctuations, $u' = u - U$, with $k_x=0$ and $k_y \in [k_f-dk_f,k_f+dk_f]$:
\begin{equation}
    a_{sj}(t) = \sum_{s\in\{-1,0,1\}} \abs{\hat u_{(0, k_f + s)(t)}}.
\end{equation}
The scales of the fluctuation amplitude  $\av{ a_{sj}}$ are indeed close to prediction \eqref{eq:square_amp_prediction}, and this for various values of $\alpha$.

The same analysis can be extended to the vortex geometry using polar coordinates. We expect that the undulations observed in SWQG arise through a similar mechanism, although their precise amplitude would depend on the extent to which small-scale generation is arrested in condensate regions.

\end{document}